\documentclass[prd,preprintnumbers,twocolumn,amsmath,nofootinbib,amssymb]{revtex4}

\usepackage{graphicx,color,dcolumn,booktabs,bm}
\usepackage{longtable,lscape}
\usepackage{makecell}
\usepackage{txfonts}
\usepackage{overpic}
\usepackage{amssymb}
\usepackage{epstopdf}
\usepackage{indentfirst}
\usepackage{feynmf}   %{feynmp}
\usepackage{slashed}  %for Feynman symbols
\usepackage{cases}
\usepackage{color}
\usepackage{float}
\usepackage{multirow}
\usepackage{ulem}
\usepackage{enumerate}
\usepackage{graphicx,color,dcolumn,booktabs,bm}
\usepackage{epsfig,dsfont,amssymb,amsmath,amsfonts,amsbsy,mathrsfs}
\graphicspath{{Figures/}} %

\usepackage{hyperref}
\hypersetup{colorlinks,citecolor=blue,anchorcolor=red,menucolor=red, linkcolor=red,filecolor=red,runcolor=red,urlcolor=blue,frenchlinks=true}

\makeatletter
\@addtoreset{equation}{section}
\makeatother

\allowdisplaybreaks

\begin{document}

\title{Five-flavor molecular pentaquarks in the $\Xi_b^{(\prime,\,*)} \bar D^{(*)}$ and $\Xi_c^{(\prime,\,*)} B^{(*)}$ systems}

\author{Fu-Lai Wang$^{1,2,3,4}$}
\email{wangfulai@lzu.edu.cn}
\author{Xiang Liu$^{1,2,3,4}$\footnote{Corresponding author}}
\email{xiangliu@lzu.edu.cn}
\affiliation{$^1$School of Physical Science and Technology, Lanzhou University, Lanzhou 730000, China\\
$^2$Lanzhou Center for Theoretical Physics, Key Laboratory of Theoretical Physics of Gansu Province, Key Laboratory of Quantum Theory and Applications of MoE, Gansu Provincial Research Center for Basic Disciplines of Quantum Physics, Lanzhou University,
Lanzhou 730000, China\\
$^3$MoE Frontiers Science Center for Rare Isotopes, Lanzhou University, Lanzhou 730000, China\\
$^4$Research Center for Hadron and CSR Physics, Lanzhou University and Institute of Modern Physics of CAS, Lanzhou 730000, China}

\begin{abstract}
The discovery of hidden-charm pentaquarks and open-flavor tetraquarks motivates the search for even more exotic hadron configurations. In this work, we investigate genuinely exotic molecular pentaquark candidates comprising five different flavors, focusing on the $\Xi_b^{(\prime,\,*)} \bar D^{(*)}$ and $\Xi_c^{(\prime,\,*)} B^{(*)}$ systems. Employing the one-boson-exchange model with the $S$-$D$ wave mixing and coupled-channel dynamics, we identify the most promising molecular pentaquark candidates comprising five different flavors. These include the $\Xi_b \bar D$, $\Xi_b^{\prime} \bar D$, $\Xi_c B$, and $\Xi_c^{\prime} B$ states with $I(J^P)=0(1/2^-)$, the $\Xi_b \bar D^{*}$, $\Xi_b^{\prime} \bar D^{*}$, $\Xi_c B^{*}$, and $\Xi_c^{\prime} B^{*}$ states with $I(J^P)=0(1/2^-,\,3/2^-)$, the $\Xi_b^{*} \bar D$ and $\Xi_c^{*} B$ states with $I(J^P)=0(3/2^-)$, as well as the $\Xi_b^{*} \bar D^{*}$ and $\Xi_c^{*} B^{*}$ states with $I(J^P)=0(1/2^-,\,3/2^-,\,5/2^-)$. Importantly, these loosely bound states exhibit pronounced spin splittings across different total angular momentum configurations after incorporating the spin-dependent interactions or the channel couplings. In addition, we identify several possible isovector molecular pentaquark candidates within the $\Xi_b^{(\prime,\,*)} \bar D^{(*)}$ and $\Xi_c^{(\prime,\,*)} B^{(*)}$ systems. Our predictions provide clear targets for experimental searches at facilities such as LHCb and Belle II, where the unique five-flavor quark configuration offers a distinctive experimental signature.
\end{abstract}
\maketitle

\section{Introduction}\label{sec1}

The exploration of the exotic hadronic states \cite{GellMann:1964nj,Zweig:1981pd} represents one of the most dynamic frontiers in the study of hadron spectroscopy. Since the turn of the century, a succession of remarkable discoveries has provided compelling evidence for the existence of the hadronic molecules \cite{Liu:2013waa,Hosaka:2016pey,Chen:2016qju,Richard:2016eis,Lebed:2016hpi,Liu:2019zoy,Brambilla:2019esw,Chen:2022asf,Olsen:2017bmm,Guo:2017jvc,Meng:2022ozq,Liu:2024uxn,Wang:2025sic,Wang:2025dur,Bai:2026atm}. Among these, the hidden-charm pentaquark structures $P_c$ discovered by the LHCb Collaboration in 2015 have played a particularly pivotal role \cite{Aaij:2015tga}. The initial observation of $P_c(4380)$ and $P_c(4450)$ in the $J/\psi p$ invariant mass spectrum of the $\Lambda_b^0 \to J/\psi p K^-$ decay was later refined with the higher-statistics data, revealing three narrow structures: $P_c(4312)$, $P_c(4440)$, and $P_c(4457)$ \cite{Aaij:2019vzc}. Their masses lie close to the $\Sigma_c \bar D$ and $\Sigma_c \bar D^{*}$ thresholds, supporting their interpretation as $S$-wave $\Sigma_c \bar D^{(*)}$ molecular states \cite{Li:2014gra,Karliner:2015ina,Wu:2010jy,Wang:2011rga,Yang:2011wz,Wu:2012md,Chen:2015loa}. Subsequently, LHCb reported two strange hidden-charm pentaquarks, $P_{cs}(4338)$ and $P_{cs}(4459)$, in the $J/\psi \Lambda$ invariant mass distribution \cite{LHCb:2020jpq,LHCb:2022ogu}, which can be explained as $\Xi_c \bar D^{(*)}$ molecular candidates \cite{Wu:2010vk,Weng:2019ynv,Wang:2019nvm,Azizi:2023foj,Wang:2022neq,Wang:2022mxy,Karliner:2022erb,Yan:2022wuz,Meng:2022wgl,Zhu:2022wpi,Chen:2022wkh,Ortega:2022uyu,Xiao:2019gjd,Anisovich:2015zqa,Chen:2016ryt,Hofmann:2005sw,Feijoo:2022rxf,Garcilazo:2022edi,Yang:2022ezl,Giachino:2022pws,Nakamura:2022jpd,Xiao:2022csb,Wang:2022gfb,Clymton:2022qlr,Chen:2022onm,Chen:2021spf,Ferretti:2021zis,Du:2021bgb,Chen:2021cfl,Hu:2021nvs,Lu:2021irg,Zou:2021sha,Wang:2021itn,Xiao:2021rgp,Zhu:2021lhd,Dong:2021juy,Wang:2020eep,Peng:2020hql,Chen:2020uif,Peng:2019wys,Chen:2020kco}. This extends the hidden-charm molecular pentaquark sector into the hidden-charm and open-strange sector.

A hadron is generally classified as exotic if its valence quark content is beyond the conventional $q\bar{q}$ or $qqq$ configurations \cite{Chen:2016qju}. In particular, hadrons composed of four different flavors are genuinely exotic and have therefore attracted intense theoretical and experimental scrutiny. The first such candidate was $X(5568)$, reported by the D\O{} Collaboration in the $B_s^0 \pi^\pm$ channel with the valence quark content $b\bar{s}u\bar{d}/b\bar{s}d\bar{u}$ \cite{D0:2016mwd}. However, subsequent searches by LHCb \cite{LHCb:2016dxl}, CDF \cite{CDF:2017dwr}, CMS \cite{CMS:2017hfy}, and ATLAS \cite{ATLAS:2018udc} failed to reproduce this signal. Fortunately, LHCb in 2020 provided important evidence for the open-flavor tetraquarks by discovering $X_0(2900)$ and $X_1(2900)$ in the $D^- K^+$ invariant mass spectrum of the $B^+ \to D^+ D^- K^+$ decay \cite{LHCb:2020bls,LHCb:2020pxc}. With the valence quark content $\bar{c} \bar{s} u d$, $X_0(2900)$ and $X_1(2900)$ are genuinely exotic hadronic candidates. Their existence was subsequently confirmed in the $B^+ \to D^{*\pm} D^{\mp} K^+$ decay \cite{LHCb:2024vfz}, with spin-parity assignments $J^P = 0^+$ for $X_0(2900)$ and $J^P = 1^-$ for $X_1(2900)$. The proximity of their masses to the $K^* \bar D^*$ and $K \bar D_1$ thresholds suggests a molecular interpretation, with $X_0(2900)$ emerging as a promising $K^* \bar D^*$ molecule and $X_1(2900)$ possibly related to a $K \bar D_1$ configuration \cite{Hu:2020mxp,Molina:2020hde,Kong:2021ohg,Wang:2021lwy,He:2020btl,Agaev:2020nrc,Chen:2020aos,Ding:2025uhh,Dai:2022htx,Chen:2023syh,Ding:2024dif,Liu:2020nil,Dong:2020rgs,Qi:2021iyv,Chen:2021tad,Wang:2025wpc}. These discoveries firmly establish the existence of the exotic hadrons comprising four different flavors and naturally motivate the search for even richer exotic hadron configurations.

As a natural extension, hadrons containing five different flavors would constitute genuinely exotic hadronic candidates. Such configurations are composed of five distinct flavors, i.e., a bottom quark, a charm quark, a strange quark, an up quark, and a down quark. However, in contrast to the extensively studied hidden-charm molecular pentaquarks, theoretical investigations of genuinely exotic molecular pentaquarks comprising five different flavors remain scarce \cite{Peng:2019wys,Shen:2022rpn,Lin:2023iww,Chen:2026cag}. Moreover, no experimental evidence for such molecular pentaquark candidates has yet been reported \cite{ParticleDataGroup:2024cfk}. This situation naturally raises a central question: what types of molecular pentaquark states with five different flavors should be prioritized in future experimental searches? Given that all observed $P_c$ \cite{Aaij:2015tga,Aaij:2019vzc} and $P_{cs}$ \cite{LHCb:2020jpq,LHCb:2022ogu} states involve a charm quark and an anti-charm quark, we focus on pentaquark systems in which one of the heavy quarks appears as an antiquark. Specifically, we consider two classes of systems: $\Xi_b^{(\prime,\,*)} \bar D^{(*)}$ and $\Xi_c^{(\prime,\,*)} B^{(*)}$. Within these, the $\Xi_b \bar D$ and $\Xi_c B$ systems can be regarded as the bottom partners of the observed $P_{cs}(4338)$ \cite{LHCb:2020jpq} within the $\Xi_c \bar D$ molecular picture, while the $\Xi_b \bar D^*$ and $\Xi_c B^*$ systems serve as the heavy-quark partners of the observed $P_{cs}(4459)$ \cite{LHCb:2022ogu} in the $\Xi_c \bar D^*$ molecular framework.

In this work, we systematically analyze the bound state properties of the $\Xi_b^{(\prime,\,*)} \bar D^{(*)}$ and $\Xi_c^{(\prime,\,*)} B^{(*)}$ systems, aiming to predict potential molecular pentaquark candidates composed of five different flavors. We begin by deriving the interaction potentials for the $\Xi_b^{(\prime,\,*)} \bar D^{(*)}$ systems within the one-boson-exchange (OBE) model, incorporating contributions from scalar, pseudoscalar, and vector meson exchanges \cite{Chen:2016qju}. In our calculation, we take into account both the $S$-$D$ wave mixing effects and coupled-channel dynamics. By solving the coupled-channel Schr\"odinger equation, we explore the formation of the loosely bound states in the $\Xi_b^{(\prime,\,*)} \bar D^{(*)}$ systems and identify the most promising molecular candidates. Subsequently, employing the heavy quark flavor symmetry, we extend our investigation to the $\Xi_c^{(\prime,\,*)} B^{(*)}$ systems. Our findings are intended to provide clear guidance for the molecular pentaquark candidates comprising five different flavors.

The present work is organized as follows. In Sec.~\ref{sec2}, we derive the interaction potentials for the $\Xi_b^{(\prime,\,*)} \bar D^{(*)}$ systems within the OBE model and analyze their bound state properties. In Sec.~\ref{sec3}, by invoking the heavy quark flavor symmetry, we extend the analysis to the $\Xi_c^{(\prime,\,*)} B^{(*)}$ systems and explore the possible formation of the loosely bound states. Finally, a brief summary and outlook are given in Sec.~\ref{sec4}.

\section{$\Xi_b^{(\prime,\,*)} \bar D^{(*)}$ systems}\label{sec2}

In this section, we present a detailed derivation of the interaction potentials for the $\Xi_b^{(\prime,\,*)} \bar D^{(*)}$ systems and analyze their bound state properties. Our aim is to identify the promising molecular pentaquark candidates composed of five different flavors.

As a key input for the study of the $\Xi_b^{(\prime,\,*)} \bar D^{(*)}$ hadronic molecular states, we first derive their interaction potentials within the OBE model, a framework widely adopted in the study of various molecular candidates \cite{Chen:2016qju}. Within this approach, the long-range part of the interaction is mediated by the pseudoscalar mesons, while the scalar and vector meson exchanges govern the medium- and short-range behavior of the interaction \cite{Chen:2016qju, Liu:2019zoy}. In the practical calculation, the momentum-space potential between the two hadrons is derived from the corresponding scattering amplitude via the Breit approximation \cite{Berestetskii:1982qgu}. Subsequently, a Fourier transformation, combined with appropriate form factors, is performed to obtain the corresponding coordinate-space potential. A detailed derivation is presented below.

A proper description of the interaction vertices is essential for constructing the scattering amplitude. The effective Lagrangians coupling the heavy hadrons to the light mesons are guided by two fundamental symmetries: the heavy quark symmetry and the chiral symmetry. Additionally incorporating the hidden local symmetry \cite{Wise:1992hn,Casalbuoni:1992gi,Casalbuoni:1996pg,Yan:1992gz,Harada:2003jx,Bando:1987br}, the couplings of the bottom baryons $\mathcal{B}_{\bar{3}}/\mathcal{B}_6^{(*)}$ and the anti-charmed mesons $\bar D^{(*)}$ to the light mesons are governed by the following effective Lagrangians \cite{Wise:1992hn,Casalbuoni:1992gi,Casalbuoni:1996pg,Yan:1992gz,Bando:1987br,Harada:2003jx,Chen:2017xat,Ding:2008gr}:
\begin{eqnarray}
\mathcal{L}_{\mathcal{B}_{\bar{3}}\mathcal{B}_{\bar{3}}\sigma}&=& l_B\langle \bar{\mathcal{B}}_{\bar{3}}\sigma\mathcal{B}_{\bar{3}}\rangle,\\
%%%
\mathcal{L}_{\mathcal{B}_{6}^{(*)}\mathcal{B}_{6}^{(*)}\sigma}&=&-l_S\langle\bar{\mathcal{B}}_6\sigma\mathcal{B}_6\rangle+l_S\langle
\bar{\mathcal{B}}_{6\mu}^{*}
\sigma\mathcal{B}_6^{*\mu}\rangle\nonumber\\
    &&-\frac{l_S}{\sqrt{3}}\langle\bar{\mathcal{B}}_{6\mu}^{*}\sigma
    \left(\gamma^{\mu}+v^{\mu}\right)\gamma^5\mathcal{B}_6\rangle+h.c.,\\
%%%
\mathcal{L}_{\mathcal{B}_6^{(*)}\mathcal{B}_6^{(*)}\mathbb{P}}&=&i\frac{g_1}{2f_{\pi}}\varepsilon^{\mu\nu\lambda\kappa}v_{\kappa}\langle\bar{\mathcal{B}}_6
\gamma_{\mu}\gamma_{\lambda}\partial_{\nu}\mathbb{P}\mathcal{B}_6\rangle\nonumber\\
    &&-i\frac{3g_1}{2f_{\pi}}\varepsilon^{\mu\nu\lambda\kappa}v_{\kappa}
    \langle\bar{\mathcal{B}}_{6\mu}^{*}\partial_{\nu}\mathbb{P}\mathcal{B}_{6\lambda}^*\rangle\nonumber\\
    &&+i\frac{\sqrt{3}g_1}{2f_{\pi}}v_{\kappa}\varepsilon^{\mu\nu\lambda\kappa}
    \langle\bar{\mathcal{B}}_{6\mu}^*\partial_{\nu}\mathbb{P}{\gamma_{\lambda}\gamma^5}
      \mathcal{B}_6\rangle+h.c.,\\
%%%
\mathcal{L}_{\mathcal{B}_{\bar{3}}\mathcal{B}_6^{(*)}\mathbb{P}} &=& -\sqrt{\frac{1}{3}}\frac{g_4}{f_{\pi}}\langle\bar{\mathcal{B}}_6\gamma^5\left(\gamma^{\mu}
+v^{\mu}\right)\partial_{\mu}\mathbb{P}\mathcal{B}_{\bar{3}}\rangle\nonumber\\
    &&-\frac{g_4}{f_{\pi}}\langle\bar{\mathcal{B}}_{6\mu}^*\partial^{\mu} \mathbb{P}\mathcal{B}_{\bar{3}}\rangle+h.c.,\\
%%%
\mathcal{L}_{\mathcal{B}_{\bar{3}}\mathcal{B}_{\bar{3}}\mathbb{V}}&=&
\frac{1}{\sqrt{2}}\beta_Bg_V\langle\bar{\mathcal{B}}_{\bar{3}}v\cdot\mathbb{V}
\mathcal{B}_{\bar{3}}\rangle,\\
%%%
\mathcal{L}_{\mathcal{B}_6^{(*)}\mathcal{B}_6^{(*)}\mathbb{V}}&=&
-\frac{\beta_Sg_V}{\sqrt{2}}\langle\bar{\mathcal{B}}_6v\cdot\mathbb{V}
\mathcal{B}_6\rangle+\frac{\beta_Sg_V}{\sqrt{2}}\langle\bar{\mathcal{B}}_{6\mu}^*v\cdot {V}\mathcal{B}_6^{*\mu}\rangle\nonumber\\
    &&-i\frac{\lambda_S g_V}{3\sqrt{2}}\langle\bar{\mathcal{B}}_6\gamma_{\mu}\gamma_{\nu}
    \left(\partial^{\mu}\mathbb{V}^{\nu}-\partial^{\nu}\mathbb{V}^{\mu}\right)
    \mathcal{B}_6\rangle\nonumber\\
    &&+i\frac{\lambda_Sg_V}{\sqrt{2}}\langle\bar{\mathcal{B}}_{6\mu}^*
    \left(\partial^{\mu}\mathbb{V}^{\nu}-\partial^{\nu}\mathbb{V}^{\mu}\right)
    \mathcal{B}_{6\nu}^*\rangle\nonumber\\
    &&-i\frac{\lambda_Sg_V}{\sqrt{6}}\langle\bar{\mathcal{B}}_{6\mu}^*
    \left(\partial^{\mu}\mathbb{V}^{\nu}-\partial^{\nu}\mathbb{V}^{\mu}\right)
    \left(\gamma_{\nu}+v_{\nu}\right)\gamma^5\mathcal{B}_6\rangle\nonumber\\
    &&-\frac{\beta_Sg_V}{\sqrt{6}}\langle\bar{\mathcal{B}}_{6\mu}^*v\cdot \mathbb{V}\left(\gamma^{\mu}+v^{\mu}\right)\gamma^5\mathcal{B}_6\rangle+h.c.,\\
%%%
\mathcal{L}_{\mathcal{B}_{\bar{3}}\mathcal{B}_6^{(*)}\mathbb{V}} &=&
-\frac{\lambda_Ig_V}{\sqrt{2}}\varepsilon^{\mu\nu\lambda\kappa}v_{\mu}\langle \bar{\mathcal{B}}_{6\nu}^*\left(\partial_{\lambda}\mathbb{V}_{\kappa}
-\partial_{\kappa}\mathbb{V}_{\lambda}\right)
          \mathcal{B}_{\bar{3}}\rangle\nonumber\\
    &&-\frac{\lambda_Ig_V}{\sqrt{6}}\varepsilon^{\mu\nu\lambda\kappa}v_{\mu}\langle \bar{\mathcal{B}}_6\gamma^5\gamma_{\nu}
        \left(\partial_{\lambda}\mathbb{V}_{\kappa}-\partial_{\kappa} \mathbb{V}_{\lambda}\right)\mathcal{B}_{\bar{3}}\rangle+h.c.,\nonumber\\\\
%%%
\mathcal{L}_{{\bar D}{\bar D}\sigma} &=&-2g_{\sigma} {\bar D}_{a} {\bar D}_a^{\dag}\sigma,\\
%%%
\mathcal{L}_{{\bar D}^{*}{\bar D}^{*}\sigma} &=&2g_{\sigma} {\bar D}_{a\mu}^* {\bar D}_a^{*\mu\dag}\sigma,\\
%%%
\mathcal{L}_{{\bar D}{\bar D}^{*}\mathbb{P}}&=&\frac{2g}{f_{\pi}}\left({\bar D}_a^{*\mu\dag}{\bar D}_b+{\bar D}_a^{\dag}{\bar D}_b^{*\mu}\right)\partial_{\mu}{\mathbb{P}}_{ab},\\
%%%
\mathcal{L}_{{\bar D}^{*}{\bar D}^{*}\mathbb{P}}&=&\frac{2ig}{f_{\pi}}v^{\alpha}\varepsilon_{\alpha\mu\nu\lambda}{\bar D}_a^{*\mu\dag}{\bar D}_b^{*\lambda}\partial^{\nu}{\mathbb{P}}_{ab},\\
%%%
\mathcal{L}_{{\bar D}{\bar D}\mathbb{V}}&=&\sqrt{2}\beta g_V {\bar D}_a {\bar D}_b^{\dag} v\cdot\mathbb{V}_{ab},\\
%%%
\mathcal{L}_{{\bar D}{\bar D}^{*}\mathbb{V}} &=&-2\sqrt{2}\lambda g_V v^{\lambda}\varepsilon_{\lambda\mu\alpha\beta}\left({\bar D}_a^{*\mu\dag}{\bar D}_b+{\bar D}_a^{\dag}{\bar D}_b^{*\mu}\right)\partial^{\alpha}\mathbb{V}^{\beta}_{ab},\nonumber\\\\
%%%
\mathcal{L}_{{\bar D}^{*}{\bar D}^{*}\mathbb{V}}&=&-\sqrt{2}\beta g_V {\bar D}_{a\mu}^* {\bar D}_b^{*\mu\dag}v\cdot\mathbb{V}_{ab}\nonumber\\
    &&-2\sqrt{2}i\lambda g_V {\bar D}_a^{*\mu\dag}{\bar D}_b^{*\nu}\left(\partial_{\mu}\mathbb{V}_{\nu}-\partial_{\nu}\mathbb{V}_{\mu}\right)_{ab}.
\end{eqnarray}
In this context, $\mathcal{B}_{\bar{3}}$ and $\mathcal{B}_6^{(*)}$ correspond to the $S$-wave bottom baryons in the $\bar{3}_F$ and $6_F$ flavor representations, respectively.

In the above effective Lagrangians, the matrix forms of $\mathcal{B}_{\bar{3}}$, $\mathcal{B}_6^{(*)}$, $\mathbb{P}$, and $\mathbb{V}_\mu$ are given by \cite{Wise:1992hn,Casalbuoni:1992gi,Casalbuoni:1996pg,Yan:1992gz,Bando:1987br,Harada:2003jx,Chen:2017xat}
\begin{eqnarray*}
\mathcal{B}_{\bar{3}} &=& \left(\begin{array}{ccc}
        0    &\Lambda_b^0      &\Xi_b^0\\
        -\Lambda_b^0  &0      &\Xi_b^-\\
        -\Xi_b^0      &-\Xi_b^-     &0
\end{array}\right),\\
\mathcal{B}_6^{(*)} &=& \left(\begin{array}{ccc}
         \Sigma_b^{{(*)}+}                  &\frac{\Sigma_b^{{(*)}0}}{\sqrt{2}}     &\frac{\Xi_b^{\prime(*)0}}{\sqrt{2}}\\
         \frac{\Sigma_b^{{(*)}0}}{\sqrt{2}}      &\Sigma_b^{{(*)}-}    &\frac{\Xi_b^{\prime(*)-}}{\sqrt{2}}\\
         \frac{\Xi_b^{\prime(*)0}}{\sqrt{2}}    &\frac{\Xi_b^{\prime(*)-}}{\sqrt{2}}      &\Omega_b^{(*)-}
\end{array}\right),\\
{\mathbb{P}} &=& {\left(\begin{array}{ccc}
       \frac{\pi^0}{\sqrt{2}}+\frac{\eta}{\sqrt{6}} &\pi^+ &K^+\\
       \pi^-       &-\frac{\pi^0}{\sqrt{2}}+\frac{\eta}{\sqrt{6}} &K^0\\
       K^-         &\bar K^0   &-\sqrt{\frac{2}{3}} \eta     \end{array}\right)},\\
{\mathbb{V}}_{\mu} &=& {\left(\begin{array}{ccc}
       \frac{\rho^0}{\sqrt{2}}+\frac{\omega}{\sqrt{2}} &\rho^+ &K^{*+}\\
       \rho^-       &-\frac{\rho^0}{\sqrt{2}}+\frac{\omega}{\sqrt{2}} &K^{*0}\\
       K^{*-}         &\bar K^{*0}   & \phi     \end{array}\right)}_{\mu}.
\end{eqnarray*}
The normalization of the heavy hadron fields is specified as follows \cite{Wise:1992hn,Casalbuoni:1992gi,Casalbuoni:1996pg,Yan:1992gz,Bando:1987br,Harada:2003jx,Chen:2017xat,Ding:2008gr}. For the bottom baryons, we have
\begin{eqnarray*}
\langle 0|\mathcal{B}|bsq\left({1}/{2}^+\right)\rangle&=& \sqrt{2m_{\mathcal{B}}}{\left(\chi_{\frac{1}{2}m},\frac{\bm{\sigma}\cdot\bm{p}}{2m_{\mathcal{B}}}\chi_{\frac{1}{2}m}\right)^T},\\
\langle 0|\mathcal{B}^{*\mu}|bsq\left({3}/{2}^+\right)\rangle &=&\sqrt{2m_{\mathcal{B}^*}}\left(\Phi_{\frac{3}{2}m}^{\mu},\frac{\bm{\sigma}\cdot\bm{p}}{2m_{\mathcal{B}^*}}\Phi_{\frac{3}{2}m}^{\mu}\right)^T,
\end{eqnarray*}
while for the anti-charmed mesons, we have
\begin{eqnarray*}
\langle 0|\bar D|\bar c{q}\left(0^-\right)\rangle&=&\sqrt{m_{\bar D}},\\
\langle 0|\bar D^{*\mu}|\bar c{q}\left(1^-\right)\rangle&=&\sqrt{m_{\bar D^*}}\epsilon^\mu.
\end{eqnarray*}
Here, $m_i$, $\bm{\sigma}$, and $\bm{p}$ denote the mass of the hadron $i$, the Pauli matrice, and the momentum of the baryon, respectively. $\chi$, $\Phi$, and $\epsilon$ represent the spin wave functions for the spin-$1/2$ baryon, the spin-$3/2$ baryon, and the spin-$1$ meson, respectively.

The coupling constants appearing in the above effective Lagrangians are typically constrained either by the theoretical considerations or, where available, by the experimental data. In this work, we adopt a set of commonly used values consistent with those in Refs. \cite{Wang:2022mxy,Chen:2017xat,Chen:2019asm,Chen:2020kco,Wang:2020bjt,Wang:2019nwt,Chen:2018pzd,Wang:2021hql,Wang:2021ajy,Wang:2021yld,Wang:2021aql,Yang:2021sue,Wang:2020dya,Wang:2023ftp,Yalikun:2023waw,Wang:2023aob,Wang:2023ael}: $l_B=-3.65$, $l_S=6.20$, $g_{\sigma}=0.76$, $g_1=0.94$, $g_4=1.06$, $g=0.59$, $f_\pi=132~\rm{MeV}$, $\beta_B g_V=-6.00$, $\beta_S g_V=12.00$, $\beta g_V=-5.25$, $\lambda_S g_V=19.20~\rm{GeV}^{-1}$, $\lambda_I g_V =-6.80~\rm{GeV}^{-1}$, and $\lambda g_V =-3.27~\rm{GeV}^{-1}$.

Having obtained the scattering amplitude $M(ab \to cd)$, the corresponding momentum-space potential can be derived within the Breit approximation \cite{Berestetskii:1982qgu}. A subsequent Fourier transformation yields the coordinate-space potential $V_E(\bm{r})$ \cite{Chen:2016qju, Liu:2019zoy}:
\begin{equation}
V_E(\bm{r}) = -\int \frac{d^3\bm{q}}{(2\pi)^3} e^{i\bm{q}\cdot\bm{r}} \frac{M(ab \to cd)}{4\sqrt{m_a m_b m_c m_d}}\,F^2(q^2),
\end{equation}
where $F(q^2)$ denotes a form factor that accounts for the finite-size effects of the interacting hadrons. A commonly used parameterization is the monopole form factor \cite{Machleidt:1987hj, Epelbaum:2008ga, Esposito:2014rxa, Chen:2016qju, Tornqvist:1993ng, Tornqvist:1993vu, Wang:2019nwt, Chen:2017jjn}, expressed as $F(q^2) = \left(\Lambda^2-m_E^2\right)/\left(\Lambda^2 - q^2\right)$, with $m_E$ being the mass of the exchanged meson and $\Lambda$ the cutoff parameter. 

In the OBE model, the cutoff parameter $\Lambda$ in the form factor is the free parameter. Owing to the absence of the experimental data on the $\Xi_b^{(\prime,\,*)} \bar D^{(*)}$ and $\Xi_c^{(\prime,\,*)} B^{(*)}$ molecular pentaquarks, $\Lambda$ cannot be uniquely fixed. In practice, the bound state solutions are searched for by scanning over a reasonable range of $\Lambda$. Following the successful OBE description of the deuteron, $\Lambda$ is generally expected to be around 1 GeV \cite{Chen:2016qju, Liu:2019zoy, Chen:2022asf}. Furthermore, in the hadronic molecular picture, the masses of the observed $P_c$ and $T_{cc}$ states can be reproduced with $\Lambda$ values near 1 GeV \cite{Chen:2016qju, Liu:2019zoy, Chen:2022asf}. As a general criterion, a loosely bound state emerging with $\Lambda \sim 1$ GeV represents a promising hadronic molecule candidate \footnote{In this regard, we emphasize that the OBE model primarily addresses the existence of hadronic molecular states, while its ability to predict precise binding energies is relatively weak. We thank the referee for highlighting this point.}.

Within the OBE model, the interaction potential between the hadrons is closely tied to the isospin and spin-orbit wave functions. We now briefly discuss both aspects. The systems under consideration involve a bottom-strange baryon $\Xi_b^{(\prime,*)}$ and an anti-charmed meson $\bar D^{(*)}$, each belonging to the isospin doublets. Consequently, the $\Xi_b^{(\prime,*)} \bar D^{(*)}$ systems can form two distinct isospin states, $I = 0$ and $1$, whose isospin wave functions are constructed via the Clebsch–Gordan coefficients. Denoting the isospin states of the bottom-strange baryons by $|\mathcal{B}_{1/2}\rangle$ and $|\mathcal{B}_{-1/2}\rangle$, and the anti-charmed meson states by $|\mathcal{M}_{1/2}\rangle$ and $|\mathcal{M}_{-1/2}\rangle$. The corresponding isospin wave functions are then given by
\begin{eqnarray*}
|I=0, I_3=0\rangle&=&\sqrt{\frac{1}{2}} \left(|\mathcal{B}_{1/2} \mathcal{M}_{-1/2}\rangle - |\mathcal{B}_{-1/2} \mathcal{M}_{1/2}\rangle \right),\\
|I=1, I_3=1\rangle&=&|\mathcal{B}_{1/2} \mathcal{M}_{1/2}\rangle,\\
|I=1, I_3=0\rangle&=&\sqrt{\frac{1}{2}} \left( |\mathcal{B}_{1/2} \mathcal{M}_{-1/2}\rangle + |\mathcal{B}_{-1/2} \mathcal{M}_{1/2}\rangle \right),\\
|I=1, I_3=-1\rangle&=&|\mathcal{B}_{-1/2} \mathcal{M}_{-1/2}\rangle.
\end{eqnarray*}
Here, $\mathcal{B}_{1/2}$ and $\mathcal{B}_{-1/2}$ denote the bottom-strange baryons with the isospin projections $I_3 = 1/2$ and $I_3 = -1/2$, respectively. Similarly, $\mathcal{M}_{1/2}$ and $\mathcal{M}_{-1/2}$ represent the anti-charmed mesons with $I_3 = 1/2$ and $I_3 = -1/2$, respectively. These expressions will be used to project the interaction potentials of the $\Xi_b^{(\prime,*)} \bar D^{(*)}$ systems onto definite isospin systems.

The ground-state $\Xi_b^{(\prime,\,*)} \bar D^{(*)}$ systems have the negative parity, arising from either the $S$-wave or the $D$-wave configurations. Among them, the $S$-wave component typically dominates, but the presence of the tensor forces can induce the $S$-$D$ mixing dynamics. The allowed spin-orbit channels for each system are listed below using the notation $|{}^{2S+1}L_J\rangle$, where $S$, $L$, and $J$ denote the total spin, orbital angular momentum, and total angular momentum, respectively.
\begin{itemize}
\item $\mathcal{B}\mathcal{M}$ system:
  \begin{equation*}
    J^P=\frac{1}{2}^-:\; \bigl|{}^2{S}_{\frac{1}{2}}\bigr\rangle.
  \end{equation*}

\item $\mathcal{B}^{*}\mathcal{M}$ system:
  \begin{equation*}
    J^P=\frac{3}{2}^-:\; \bigl|{}^4{S}_{\frac{3}{2}}\bigr\rangle,\; \bigl|{}^4{D}_{\frac{3}{2}}\bigr\rangle.
  \end{equation*}

\item $\mathcal{B}\mathcal{M}^{*}$ system:
  \begin{align*}
    J^P=\frac{1}{2}^- &:\; \bigl|{}^2{S}_{\frac{1}{2}}\bigr\rangle,\; \bigl|{}^4{D}_{\frac{1}{2}}\bigr\rangle,\\
    J^P=\frac{3}{2}^- &:\; \bigl|{}^4{S}_{\frac{3}{2}}\bigr\rangle,\; \bigl|{}^2{D}_{\frac{3}{2}}\bigr\rangle,\; \bigl|{}^4{D}_{\frac{3}{2}}\bigr\rangle.
  \end{align*}

\item $\mathcal{B}^{*}\mathcal{M}^{*}$ system:
  \begin{align*}
    J^P=\frac{1}{2}^- &:\; \bigl|{}^2{S}_{\frac{1}{2}}\bigr\rangle,\; \bigl|{}^4{D}_{\frac{1}{2}}\bigr\rangle,\; \bigl|{}^6{D}_{\frac{1}{2}}\bigr\rangle,\\
    J^P=\frac{3}{2}^- &:\; \bigl|{}^4{S}_{\frac{3}{2}}\bigr\rangle,\; \bigl|{}^2{D}_{\frac{3}{2}}\bigr\rangle,\; \bigl|{}^4{D}_{\frac{3}{2}}\bigr\rangle,\; \bigl|{}^6{D}_{\frac{3}{2}}\bigr\rangle,\\
    J^P=\frac{5}{2}^- &:\; \bigl|{}^6{S}_{\frac{5}{2}}\bigr\rangle,\; \bigl|{}^2{D}_{\frac{5}{2}}\bigr\rangle,\; \bigl|{}^4{D}_{\frac{5}{2}}\bigr\rangle,\; \bigl|{}^6{D}_{\frac{5}{2}}\bigr\rangle.
  \end{align*}
\end{itemize}
The corresponding spin-orbit wave functions for the $\Xi_b^{(\prime,*)}\bar D^{(*)}$ systems are constructed as
\begin{align}
\mathcal{B}\mathcal{M}&:\;  \sum_{m,\,m_L} C^{JM}_{\frac{1}{2}m,Lm_L}\; \chi_{\frac{1}{2}m}\, Y_{L m_L}, \nonumber\\
\mathcal{B}^{*}\mathcal{M}&:\; \sum_{m,\,m_L} C^{JM}_{\frac{3}{2}m,Lm_L}\; \Phi_{\frac{3}{2}m}\, Y_{L m_L}, \nonumber\\
\mathcal{B}\mathcal{M}^{*}&:\; \sum_{m,\,m',\,m_S,\,m_L} C^{Sm_S}_{\frac{1}{2}m,1m'}\, C^{JM}_{Sm_S,Lm_L}\; \chi_{\frac{1}{2}m}\,\epsilon^{\mu}_{m'}\, Y_{L m_L}, \nonumber\\
\mathcal{B}^{*}\mathcal{M}^{*}&:\; \sum_{m,\,m',\,m_S,\,m_L}C^{Sm_S}_{\frac{3}{2}m,1m'}\, C^{JM}_{Sm_S,Lm_L}\; \Phi_{\frac{3}{2}m}^{\mu}\,\epsilon^{\nu}_{m'}\, Y_{L m_L}. \nonumber
\end{align}
Here, $Y_{L m_L}$ denotes the spherical harmonic function. Within the framework established above, the interaction potentials for the $\Xi_b^{(\prime,\,*)} \bar D^{(*)}$ systems can be derived. Their explicit forms can refer to Refs.~\cite{Chen:2020kco, Wang:2022mxy, Chen:2022onm}.

Based on the interaction potentials of the $\Xi_b^{(\prime,\,*)} \bar D^{(*)}$ systems, we now investigate their bound state properties. By solving the coupled-channel Schr\"odinger equation, we search for the loosely bound states. For each bound state solution, we compute the binding energy $E$ and the spatial wave functions $\psi_i(r)$ of the individual channel $i$. From these wave functions, we further determine the root-mean-square (RMS) radius $r_{\rm RMS}$ and the channel probability $P_i$ using the following relations:
\begin{eqnarray}
r_{\rm RMS} &=& \sqrt{\int \sum_i \psi_i^{\dagger}(r) \psi_i(r) \, r^4 \, dr},\\
P_i &=& \int \psi_i^{\dagger}(r) \psi_i(r) \, r^2 \, dr,
\end{eqnarray}
where the wave functions satisfy the normalization condition $\int \sum_i \psi_i^{\dagger}(r) \psi_i(r) \, r^2 \, dr = 1$. The binding energy $E$, the root-mean-square radius $r_{\rm RMS}$, and the channel probabilities $P_i$ collectively characterize the bound state properties of the $\Xi_b^{(\prime,\,*)} \bar D^{(*)}$ systems. Notably, $r_{\rm RMS}$ reflects the spatial extension of the molecular state. As the cutoff parameter $\Lambda$ increases, the binding energy deepens and the root-mean-square radius decreases, consistent with the inverse relation between $E$ and $r_{\rm RMS}^2$ \cite{Chen:2017xat}. A promising hadronic molecule candidate is typically expected to have the root-mean-square radius larger than the size of all the included component hadrons.

In the following, we first present the single-channel analysis of the bound state properties for the $\Xi_b^{(\prime,\,*)} \bar D^{(*)}$ systems, incorporating the $S$-$D$ wave mixing effects. Subsequently, we briefly discuss the influence of the coupled-channel dynamics on the bound state properties for the $\Xi_b^{(\prime,\,*)} \bar D^{(*)}$ systems. Our results provide critical insights into the bound state properties of the $\Xi_b^{(\prime,\,*)} \bar D^{(*)}$ systems and highlight promising molecular pentaquark candidates for future experimental searches.

\renewcommand\tabcolsep{0.46cm}
\renewcommand{\arraystretch}{1.50}
\begin{table}[!htbp]
\caption{Bound state properties for the $\Xi_b \bar D$ and  $\Xi_b^{\prime} \bar D$ systems.}\label{massspectra1}
\begin{tabular}{c|ccc}\toprule[1pt]\toprule[1pt]
States&$\Lambda\,{\rm (GeV)}$ &$E\,{\rm (MeV)}$  &$r_{\rm RMS}\,{\rm (fm)}$\\\midrule[1.0pt]
\multirow{3}{*}{$\Xi_b \bar D[0(\frac{1}{2}^{-})]$}
&$1.25$&$-0.36$&$4.31$\\
&$1.40$&$-4.86$&$1.47$\\
&$1.54$&$-12.64$&$1.00$\\\midrule[1.0pt]
\multirow{3}{*}{$\Xi_b^{\prime} \bar D[0(\frac{1}{2}^{-})]$}
&$1.28$&$-0.33$&$4.44$\\
&$1.43$&$-4.40$&$1.53$\\
&$1.58$&$-12.05$&$1.00$\\
\bottomrule[1pt]\bottomrule[1pt]
\end{tabular}
\end{table}

The $\Xi_b \bar D$ and $\Xi_b^{\prime} \bar D$ systems are composed of a spin-$1/2$ baryon and a spin-$0$ meson, which uniquely determines the total spin-parity quantum number as $J^P = 1/2^-$, with only the $S$-wave contribution. Table~\ref{massspectra1} summarizes the bound state solutions obtained for these systems.

For the $\Xi_b \bar D$ state with $I(J^P)=0(1/2^-)$, a loosely bound state first appears at a cutoff $\Lambda = 1.25$ GeV, characterized by a binding energy $E = -0.36$ MeV and a root-mean-square radius $r_{\rm RMS} = 4.31$ fm. As $\Lambda$ increases to $1.54$ GeV, the binding deepens to $E = -12.64$ MeV and the radius contracts to $1.00$ fm. A similar pattern is observed for the $\Xi_b^{\prime} \bar D$ state with $I(J^P)=0(1/2^-)$: a loosely bound state emerges at $\Lambda = 1.28$ GeV with $E = -0.33$ MeV and $r_{\rm RMS} = 4.44$ fm, and at $\Lambda = 1.58$ GeV, $E = -12.05$ MeV and $r_{\rm RMS} = 1.00$ fm. This behavior demonstrates that the $\Xi_b \bar D$ and $\Xi_b^{\prime} \bar D$ states with $I(J^P)=0(1/2^-)$ can form the loosely bound states within a reasonable range of the cutoffs. Consequently, the  $\Xi_b \bar D$ and $\Xi_b^{\prime} \bar D$ states with $I(J^P)=0(1/2^-)$ emerge as the most promising molecular pentaquark candidates comprising five different flavors and merit prioritized attention in future experimental searches.

\renewcommand\tabcolsep{0.15cm}
\renewcommand{\arraystretch}{1.50}
\begin{table}[!htbp]
\caption{Bound state properties for the $\Xi_b \bar D^{*}$ and $\Xi_b^{*} \bar D$ systems.}\label{massspectra2}
\begin{tabular}{c|cccc}\toprule[1pt]\toprule[1pt]
States&$\Lambda\,{\rm (GeV)}$ &$E\,{\rm (MeV)}$  &$r_{\rm RMS}\,{\rm (fm)}$& {Probability (\%)}\\\midrule[1.0pt]
\multirow{3}{*}{$\Xi_b \bar D^{*}[0(\frac{1}{2}^{-})]$}
&$1.22$&$-0.30$&$4.48$&{100.00} (${}^2{S}_{\frac{1}{2}}$)\\
&$1.36$&$-4.46$&$1.50$&{100.00} (${}^2{S}_{\frac{1}{2}}$)\\
&$1.50$&$-12.42$&$0.99$&{100.00} (${}^2{S}_{\frac{1}{2}}$)\\\hline
\multirow{3}{*}{$\Xi_b \bar D^{*}[0(\frac{3}{2}^{-})]$}
&$1.22$&$-0.30$&$4.48$&{100.00} (${}^4{S}_{\frac{3}{2}}$)\\
&$1.36$&$-4.46$&$1.50$&{100.00} (${}^4{S}_{\frac{3}{2}}$)\\
&$1.50$&$-12.42$&$0.99$&{100.00} (${}^4{S}_{\frac{3}{2}}$)\\\midrule[1.0pt]
\multirow{3}{*}{$\Xi_b^{*} \bar D[0(\frac{3}{2}^{-})]$}
&$1.28$&$-0.33$&$4.42$&{100.00} (${}^4{S}_{\frac{3}{2}}$)\\
&$1.44$&$-4.83$&$1.47$&{100.00} (${}^4{S}_{\frac{3}{2}}$)\\
&$1.59$&$-12.70$&$0.99$&{100.00} (${}^4{S}_{\frac{3}{2}}$)\\
\bottomrule[1pt]\bottomrule[1pt]
\end{tabular}
\end{table}

Although the $D$-wave channels are kinematically accessible for the $\Xi_b \bar D^{*}$ and $\Xi_b^{*} \bar D$ systems, the tensor force is absent from the corresponding interaction potentials. Consequently, the $D$-wave components do not contribute, and the bound state solutions are determined purely by the $S$-wave interactions. Table~\ref{massspectra2} lists the resulting bound state solutions for the $\Xi_b \bar D^{*}$ and $\Xi_b^{*} \bar D$ systems.

For the $\Xi_b \bar D^{*}$ states with $I(J^P)=0(1/2^-)$ and $0(3/2^-)$, the loosely bound states appear at $\Lambda = 1.22$ GeV, with binding energy $E = -0.30$ MeV and root-mean-square radius $r_{\rm RMS} = 4.48$ fm, arising purely from the ${}^2S_{1/2}$ and ${}^4S_{3/2}$ partial waves, respectively. As the cutoff increases to $1.50$ GeV, the binding deepens to $E = -12.42$ MeV and the radius contracts to $0.99$ fm. A notable feature is the exact degeneracy observed between the $\Xi_b \bar D^{*}$ bound states with $I(J^P)=0(1/2^-)$ and $0(3/2^-)$ for a given cutoff value: the binding energies and the root-mean-square radii are identical for both spin configurations. This degeneracy arises because, within the single-channel framework, the effective potential does not distinguish between different total spin configurations when the spin-dependent term is absent from the interaction. This behavior mirrors our previous findings for the $\Xi_c \bar D^{*}$ system in Ref.~\cite{Wang:2022mxy}. The $\Xi_b \bar D^{*}$ loosely bound states with $I(J^P)=0(1/2^-)$ and $0(3/2^-)$ can be formed within a reasonable range of the cutoffs and therefore emerge as the most promising molecular pentaquark candidates comprising five different flavors.

For the $\Xi_b^{*} \bar D$ system, only the $J^P = 3/2^-$ configuration is allowed. The $\Xi_b^{*} \bar D$ state with $I(J^P)=0(3/2^-)$ forms a loosely bound state at the cutoff $\Lambda = 1.28$ GeV, with binding energy $E = -0.33$ MeV and root-mean-square radius $r_{\rm RMS} = 4.42$ fm, arising purely from the ${}^4S_{3/2}$ partial wave. Increasing $\Lambda$ to $1.59$ GeV yields $E = -12.70$ MeV and $r_{\rm RMS} = 0.99$ fm. These results indicate that the $\Xi_b^{*} \bar D$ state with $I(J^P)=0(3/2^-)$ also emerges as a most promising molecular pentaquark candidate comprising five different flavors.

\renewcommand\tabcolsep{0.20cm}
\renewcommand{\arraystretch}{1.50}
\begin{table}[!htbp]
\caption{Bound state properties for the $\Xi_b^{\prime} \bar D^{*}$ and $\Xi_b^{*} \bar D^{*}$ systems, with $\Lambda$, $E$, and $r_{\rm RMS}$ given in GeV, MeV, and fm, respectively. The dominant channel for each bound state is denoted in bold.}\label{massspectra3}
\begin{tabular}{c|cccc}\toprule[1pt]\toprule[1pt]
\multirow{4}{*}{$\Xi_b^{\prime} \bar D^{*}[0(\frac{1}{2}^{-})]$}
&$\Lambda$ &$E$  &$r_{\rm RMS}$ &P(${}^2{S}_{\frac{1}{2}}/{}^4{D}_{\frac{1}{2}})$\\
&0.83&$-0.44$ &3.83&\textbf{99.47}/0.53\\
&0.89&$-4.39$ &1.44&\textbf{99.25}/0.75\\
&0.94&$-11.83$ &0.96&\textbf{99.29}/0.71\\\hline
\multirow{4}{*}{$\Xi_b^{\prime} \bar D^{*}[0(\frac{3}{2}^{-})]$}
&$\Lambda$ &$E$  &$r_{\rm RMS}$ &P(${}^4{S}_{\frac{3}{2}}/{}^2{D}_{\frac{3}{2}}/{}^4{D}_{\frac{3}{2}})$\\
&1.31&$-0.31$ &4.65&\textbf{97.84}/0.33/1.83\\
&1.62&$-4.73$ &1.63&\textbf{94.65}/0.79/4.56\\
&1.92&$-12.12$ &1.15&\textbf{93.21}/1.03/5.76\\\midrule[1.0pt]
\multirow{4}{*}{$\Xi_b^{*} \bar D^{*}[0(\frac{1}{2}^{-})]$}
&$\Lambda$ &$E$  &$r_{\rm RMS}$ &P(${}^2{S}_{\frac{1}{2}}/{}^4{D}_{\frac{1}{2}}/{}^6{D}_{\frac{1}{2}})$\\
&0.79&$-1.99$ &2.04&\textbf{98.77}/0.76/0.48\\
&0.84&$-6.24$ &1.27&\textbf{98.66}/0.84/0.51\\
&0.88&$-12.49$ &0.96&\textbf{98.71}/0.81/0.48\\\hline
\multirow{4}{*}{$\Xi_b^{*} \bar D^{*}[0(\frac{3}{2}^{-})]$}
&$\Lambda$ &$E$  &$r_{\rm RMS}$ &P(${}^4{S}_{\frac{3}{2}}/{}^2{D}_{\frac{3}{2}}/{}^4{D}_{\frac{3}{2}}/{}^6{D}_{\frac{3}{2}})$\\
&0.94&$-0.43$ &3.95&\textbf{98.47}/0.40/1.01/0.12\\
&1.01&$-4.18$ &1.53&\textbf{97.52}/0.66/1.64/0.18\\
&1.08&$-12.40$ &0.98&\textbf{97.42}/0.71/1.69/0.18\\\hline
\multirow{4}{*}{$\Xi_b^{*} \bar D^{*}[0(\frac{5}{2}^{-})]$}
&$\Lambda$ &$E$  &$r_{\rm RMS}$ &P(${}^6{S}_{\frac{5}{2}}/{}^2{D}_{\frac{5}{2}}/{}^4{D}_{\frac{5}{2}}/{}^6{D}_{\frac{5}{2}})$\\
&1.31&$-0.30$ &4.75&\textbf{97.15}/0.14/0.09/2.63\\
&1.62&$-4.52$ &1.70&\textbf{92.54}/0.31/0.22/6.94\\
&1.92&$-12.10$ &1.19&\textbf{90.30}/0.39/0.27/9.04\\\hline
\multirow{4}{*}{$\Xi_b^{*} \bar D^{*}[1(\frac{5}{2}^{-})]$}
&$\Lambda$ &$E$  &$r_{\rm RMS}$ &P(${}^6{S}_{\frac{5}{2}}/{}^2{D}_{\frac{5}{2}}/{}^4{D}_{\frac{5}{2}}/{}^6{D}_{\frac{5}{2}})$\\
&1.70&$-0.32$ &3.97&\textbf{99.48}/0.01/$\mathcal{O}(0)$/0.10\\
&1.73&$-3.69$ &1.26&\textbf{99.81}/0.01/0.01/0.18\\
&1.75&$-7.48$ &0.89&\textbf{99.76}/0.01/0.01/0.22\\
\bottomrule[1pt]\bottomrule[1pt]
\end{tabular}
\end{table}

The $\Xi_b^{\prime} \bar D^{*}$ and $\Xi_b^{*} \bar D^{*}$ systems involve both baryon and meson having spin, leading to a variety of accessible $J^P$ quantum numbers. Moreover, the tensor force is present in their interaction potentials, giving rise to the $S$-$D$ wave mixing effects. The bound state properties for these systems are summarized in Table~\ref{massspectra3}.

For the $\Xi_b^{\prime} \bar D^{*}$ state with $I(J^P)=0(1/2^-)$, a loosely bound state appears at a relatively low cutoff $\Lambda = 0.83$ GeV, with binding energy $E = -0.44$ MeV and root-mean-square radius $r_{\rm RMS} = 3.83$ fm. The wave function is predominantly composed of the ${}^2S_{1/2}$ partial wave, with a small admixture of the ${}^4D_{1/2}$ component. As the cutoff $\Lambda$ increases to $0.94$ GeV, the binding deepens to $-11.83$ MeV and the radius contracts to $0.96$ fm. Thus, the $\Xi_b^{\prime} \bar D^{*}$ state with $I(J^P)=0(1/2^-)$ can form a loosely bound state within a reasonable cutoff range, establishing it as a strong molecular candidate.
For the $\Xi_b^{\prime} \bar D^{*}$ state with $I(J^P)=0(3/2^-)$, a loosely bound state first appears at a larger cutoff $\Lambda = 1.31$ GeV, with $E = -0.31$ MeV and $r_{\rm RMS} = 4.65$ fm, and the wave function is dominated by the ${}^4S_{3/2}$ partial wave. At $\Lambda = 1.92$ GeV, we obtain $E = -12.12$ MeV and $r_{\rm RMS} = 1.15$ fm. The larger cutoff required for binding indicates a weaker attraction compared to the $\Xi_b^{\prime} \bar D^{*}$ bound state with $I(J^P)=0(1/2^-)$, demonstrating a clear spin-dependent splitting. Consequently, the $\Xi_b^{\prime} \bar D^{*}$ bound states with $I(J^P)=0(1/2^-)$ and $0(3/2^-)$ are not degenerate, and the spin-spin interaction splits them significantly. Both bound states can be formed within a reasonable ranges of the cutoff and therefore emerge as the most promising molecular pentaquark candidates comprising five different flavors.

For the $\Xi_b^{*} \bar D^{*}$ system, four distinct loosely bound states can be formed within a cutoff range of $0.8$ to $2.0$ GeV, corresponding to the quantum numbers $I(J^P)=0(1/2^-)$, $0(3/2^-)$, $0(5/2^-)$, and $1(5/2^-)$. In the following, we discuss each of these bound states in detail.
For the $\Xi_b^* \bar D^{*}$ state with $I(J^P)=0(1/2^-)$, there exists the loosely bound state at a relatively low cutoff $\Lambda = 0.79$ GeV, with binding energy $E = -1.99$ MeV and root-mean-square radius $r_{\rm RMS} = 2.04$ fm. The wave function is dominated by the ${}^2S_{1/2}$ component, with tiny admixtures from the ${}^4D_{1/2}$ and ${}^6D_{1/2}$ partial waves. As $\Lambda$ increases to $0.88$ GeV, the binding deepens to $E = -12.49$ MeV and the radius contracts to $0.96$ fm, while the $D$-wave fractions remain below $1\%$. The appearance of this bound state at such low cutoff indicates strong attraction, establishing it as an excellent candidate for a molecular pentaquark comprising five different flavors.
For the $\Xi_b^* \bar D^{*}$ state with $I(J^P)=0(3/2^-)$, a loosely bound state appears at $\Lambda = 0.94$ GeV with $E = -0.43$ MeV and $r_{\rm RMS} = 3.95$ fm. The main component is the ${}^4S_{3/2}$ partial wave, with the $D$-wave contributions from the ${}^2D_{3/2}$, ${}^4D_{3/2}$, and ${}^6D_{3/2}$ channels summing to approximately $1.53\%$. Increasing $\Lambda$ to $1.08$ GeV yields $E = -12.40$ MeV and $r_{\rm RMS} = 0.98$ fm. This state also binds at reasonable cutoff and represents a viable molecular pentaquark candidate with five different flavors.
For the $\Xi_b^* \bar D^{*}$ state with $I(J^P)=0(5/2^-)$, a loosely bound state first appears at $\Lambda = 1.31$ GeV with $E = -0.30$ MeV and $r_{\rm RMS} = 4.75$ fm. The ${}^6S_{5/2}$ wave dominates, while the $D$-wave components (${}^2D_{5/2}$, ${}^4D_{5/2}$, and ${}^6D_{5/2}$) collectively amount to $2.86\%$. As $\Lambda$ increases to $1.92$ GeV, $E = -12.10$ MeV and $r_{\rm RMS} = 1.19$ fm. Thus, the $\Xi_b^* \bar D^{*}$ state with $I(J^P)=0(5/2^-)$ forms a loosely bound state within a reasonable cutoff range and emerges as a most promising molecular pentaquark candidate comprising five different flavors.
For the $\Xi_b^* \bar D^{*}$ state with $I(J^P)=1(5/2^-)$, a loosely bound state appears at $\Lambda = 1.70$ GeV with $E = -0.32$ MeV and $r_{\rm RMS} = 3.97$ fm, dominated by the ${}^6S_{5/2}$ wave with negligible $D$-wave admixtures. Increasing $\Lambda$ to $1.73$ GeV yields $E = -3.69$ MeV and $r_{\rm RMS} = 1.26$ fm. Although the required cutoff is higher than for its isoscalar partners, it remains within an acceptable range. Consequently, the $\Xi_b^* \bar D^{*}$ state with $I(J^P)=1(5/2^-)$ can be regarded as a possible isovector molecular pentaquark candidate.

The non-degeneracy among the $\Xi_b^{\prime} \bar D^{*}$ and $\Xi_b^{*} \bar D^{*}$ bound states is clearly demonstrated for a given cutoff value: each bound state with distinct $I(J^P)$ exhibits a unique dependence on the cutoffs and the bound state properties. In particular, the $I(J^P)=0(1/2^-)$ state is the most deeply bound for a given $\Lambda$, whereas the $I(J^P)=0(3/2^-,\,5/2^-)$ states require the largest cutoff to bind. This pattern reflects the spin-dependent nature of the interaction potentials, which leads to significant multiplet splitting. Such splittings will be essential for distinguishing these molecular pentaquark candidates comprising five different flavors in future experimental searches, analogous to the role played by the fine structure in the identification of the $P_c(4440)$ and $P_c(4457)$ \cite{Aaij:2019vzc} in the $\Sigma_c \bar D^*$ molecular framework.

We now proceed to perform the coupled-channel calculations for the $\Xi_b^{(\prime,\,*)} \bar D^{(*)}$ systems, treating the isoscalar and isovector sectors separately. In the coupled-channel analysis, the channel probabilities $P_i$ provide valuable insight into the nature of a hadronic molecular state, in addition to the binding energy $E$ and the root-mean-square radius $r_{\rm RMS}$. If the coupled system is not dominated by the channel with the lowest mass threshold among the included channels, the resulting $r_{\rm RMS}$ typically turns out to be too small, and such a system is unlikely to qualify as a genuine hadronic molecule (see Ref. \cite{Chen:2017xat} for more details). For each coupled system, we retain only those solutions in which the dominant channel corresponds to the lowest threshold among the coupled channels, thereby ensuring that the resulting bound state is consistent with the picture of a loosely bound state \cite{Chen:2017xat}.

\renewcommand\tabcolsep{0.31cm}
\renewcommand{\arraystretch}{1.50}
\begin{table}[!htbp]
\caption{Bound state properties for the isoscalar $\Xi_b^{(\prime,\,*)} \bar D^{(*)}$ systems obtained from the coupled-channel calculations, with $\Lambda$, $E$, and $r_{\rm RMS}$ given in GeV, MeV, and fm, respectively. The dominant channel for each bound state is denoted in bold.}\label{massspectra4}
\begin{tabular}{cccc}\toprule[1pt]\toprule[1pt]
\multicolumn{4}{c}{$\Xi_b \bar D/\Xi_b^{\prime} \bar D/\Xi_b \bar D^*/\Xi_b^{\prime} \bar D^*/\Xi_b^{*} \bar D^*$ coupled system with $0(1/2^-)$}\\\hline
$\Lambda$ &$E$  &$r_{\rm RMS}$& P($\Xi_b \bar D/\Xi_b^{\prime} \bar D/\Xi_b \bar D^*/\Xi_b^{\prime} \bar D^*/\Xi_b^{*} \bar D^*$)\\
0.99&$-0.32$ &4.39&\textbf{97.96}/$\mathcal{O}(0)$/$\mathcal{O}(0)$/0.77/1.28\\
1.03&$-4.08$ &1.46&\textbf{90.44}/0.02/0.02/3.77/5.75\\
1.06&$-10.73$ &0.93&\textbf{81.18}/0.08/0.12/8.04/10.57\\\midrule[1.0pt]
\multicolumn{4}{c}{$\Xi_b \bar D^*/\Xi_b^{\prime} \bar D^*/\Xi_b^{*} \bar D^*$ coupled system with $0(1/2^-)$}\\\hline
$\Lambda$ &$E$  &$r_{\rm RMS}$& P($\Xi_b \bar D^*/\Xi_b^{\prime} \bar D^*/\Xi_b^{*} \bar D^*$)\\
0.94&$-0.20$ &4.88&\textbf{97.75}/1.64/0.61\\
0.97&$-3.53$ &1.51&\textbf{89.77}/7.55/2.68\\
1.00&$-11.66$ &0.87&\textbf{80.27}/14.18/4.96\\\midrule[1.0pt]
\multicolumn{4}{c}{$\Xi_b \bar D^*/\Xi_b^* \bar D/\Xi_b^{\prime} \bar D^*/\Xi_b^{*} \bar D^*$ coupled system with $0(3/2^-)$}\\\hline
$\Lambda$ &$E$  &$r_{\rm RMS}$& P($\Xi_b \bar D^*/\Xi_b^* \bar D/\Xi_b^{\prime} \bar D^*/\Xi_b^{*} \bar D^*$)\\
0.81&$-0.40$ &3.84&\textbf{89.64}/8.38/0.08/1.90\\
0.85&$-4.17$ &1.34&\textbf{73.01}/20.66/0.26/6.07\\
0.89&$-11.77$ &0.87&\textbf{63.11}/25.98/0.51/10.41\\\midrule[1.0pt]
\multicolumn{4}{c}{$\Xi_b^* \bar D/\Xi_b^{\prime} \bar D^*/\Xi_b^{*} \bar D^*$ coupled system with $0(3/2^-)$}\\\hline
$\Lambda$ &$E$  &$r_{\rm RMS}$& P($\Xi_b^* \bar D/\Xi_b^{\prime} \bar D^*/\Xi_b^{*} \bar D^*$)\\
1.06&$-0.41$ &4.03&\textbf{97.64}/0.01/2.36\\
1.12&$-4.68$ &1.37&\textbf{89.00}/0.13/10.87\\
1.17&$-12.96$ &0.85&\textbf{76.69}/0.72/22.59\\\midrule[1.0pt]
\multicolumn{4}{c}{$\Xi_b^{\prime} \bar D^*/\Xi_b^{*} \bar D^*$ coupled system with $0(1/2^-)$}\\\hline
$\Lambda$ &$E$  &$r_{\rm RMS}$& P($\Xi_b^{\prime} \bar D^*/\Xi_b^{*} \bar D^*$)\\
0.81&$-1.09$ &2.50&\textbf{91.50}/8.50\\
0.85&$-5.56$ &1.18&\textbf{77.45}/22.55\\
0.88&$-11.89$ &0.86&\textbf{66.79}/33.21\\
\bottomrule[1pt]\bottomrule[1pt]
\end{tabular}
\end{table}

Table~\ref{massspectra4} summarizes the bound state solutions for the isoscalar $\Xi_b^{(\prime,\,*)} \bar D^{(*)}$ systems obtained from the coupled-channel calculations. In these systems, the inclusion of the coupled-channel effects modifies the bound state properties but does not introduce additional loosely bound states beyond those already identified in the single-channel analysis. The most significant finding is the lifting of the degeneracy previously observed in the single-channel analysis between the $\Xi_b \bar D^{*}$ bound states with $I(J^P)=0(1/2^-)$ and $0(3/2^-)$. We now discuss these findings in detail.

For the $\Xi_b \bar D/\Xi_b^{\prime} \bar D/\Xi_b \bar D^*/\Xi_b^{\prime} \bar D^*/\Xi_b^{*} \bar D^*$ coupled system with $I(J^P)=0(1/2^-)$, a loosely bound state emerges at $\Lambda = 0.99$ GeV, with binding energy $E = -0.32$ MeV and root-mean-square radius $r_{\rm RMS} = 4.39$ fm. The $\Xi_b \bar D$ channel dominates the wave function, with small admixtures from the $\Xi_b^{\prime} \bar D^*$ and $\Xi_b \bar D^*$ components. Increasing the cutoff to $\Lambda = 1.06$ GeV deepens the binding to $E = -10.73$ MeV and reduces the radius to $r_{\rm RMS} = 0.93$ fm, while the $\Xi_b \bar D$ channel remains dominant. Compared to the single-channel $\Xi_b \bar D$ bound state with identical quantum numbers, the coupled-channel calculation yields slightly deeper binding at the same cutoff, indicating that the channel couplings provide a moderate attractive enhancement. This feature persists across the other isoscalar $\Xi_b^{(\prime,\,*)} \bar D^{(*)}$ systems discussed.

\begin{figure}[htbp]
\centering
\includegraphics[width=8.6cm]{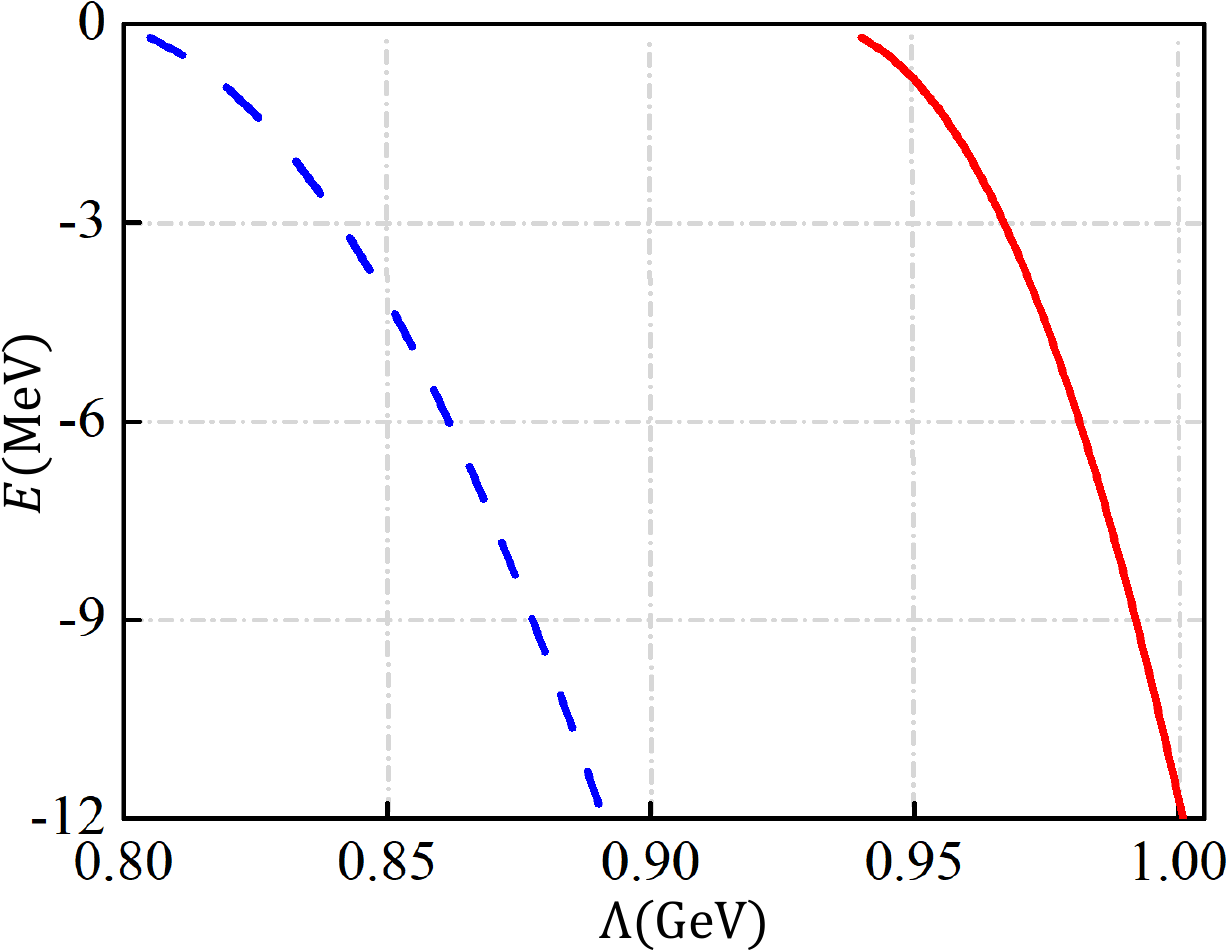}
\caption{Relationship between the binding energy and the cutoff in the $\Xi_b \bar D^* / \Xi_b' \bar D^* / \Xi_b^* \bar D^*$ coupled system with $I(J^P)=0(1/2^-)$ and the $\Xi_b \bar D^* / \Xi_b^* \bar D / \Xi_b' \bar D^* / \Xi_b^* \bar D^*$ coupled system with $I(J^P)=0(3/2^-)$. These two cases are represented by the red solid line and the blue dashed line, respectively.}\label{degeneracy}
\end{figure}

The most significant finding from the coupled-channel calculations is the removal of the degeneracy previously observed in the single-channel analysis between the $\Xi_b \bar D^{*}$ bound states with $I(J^P)=0(1/2^-)$ and $0(3/2^-)$. For the $\Xi_b \bar D^*/\Xi_b^{\prime} \bar D^*/\Xi_b^{*} \bar D^*$ coupled system with $I(J^P)=0(1/2^-)$, a loosely bound state emerges at $\Lambda = 0.94$ GeV with $E = -0.20$ MeV and $r_{\rm RMS} = 4.88$ fm, dominated by the $\Xi_b \bar D^*$ channel. As $\Lambda$ increases to $1.00$ GeV, the binding deepens to $E = -11.66$ MeV and the radius contracts to $0.87$ fm, with the $\Xi_b \bar D^*$ channel remaining the dominant component. For the $\Xi_b \bar D^*/\Xi_b^* \bar D/\Xi_b^{\prime} \bar D^*/\Xi_b^{*} \bar D^*$ coupled system with $I(J^P)=0(3/2^-)$, a loosely bound state appears at $\Lambda = 0.81$ GeV with $E = -0.40$ MeV and $r_{\rm RMS} = 3.84$ fm. Here, the dominant channel is the $\Xi_b \bar D^*$ channel, followed by the $\Xi_b^* \bar D$ channel. At $\Lambda = 0.89$ GeV, $E = -11.77$ MeV and $r_{\rm RMS} = 0.87$ fm, with the $\Xi_b \bar D^*$ channel still dominant but accompanied by a significant $\Xi_b^* \bar D$ component. As illustrated in Fig. \ref{degeneracy}, the binding energies of the $I(J^P)=0(1/2^-)$ and $0(3/2^-)$ $\Xi_b \bar D^{*}$ coupled systems exhibit distinct dependencies on the cutoff parameter, thereby completely removing the degeneracy observed in the single-channel $\Xi_b \bar D^{*}$ bound states. This breaking of degeneracy arises from the fact that the two $J^P$ states couple to different sets of nearby hadron channels, yielding distinct effective interactions in each partial wave. A similar phenomenon was identified in our previous study of the $\Xi_c \bar D^{*}$ system \cite{Wang:2022mxy}, emphasizing the critical role of the coupled-channel dynamics in determining the mass spectrum of the hadronic molecules.

\renewcommand\tabcolsep{0.40cm}
\renewcommand{\arraystretch}{1.50}
\begin{table}[!htbp]
\caption{Bound state properties for the isovector $\Xi_b^{(\prime,\,*)} \bar D^{(*)}$ systems obtained from the coupled-channel calculations, with $\Lambda$, $E$, and $r_{\rm RMS}$ given in GeV, MeV, and fm, respectively. The dominant channel for each bound state is denoted in bold.}\label{massspectra5}
\begin{tabular}{cccc}\toprule[1pt]\toprule[1pt]
\multicolumn{4}{c}{$\Xi_b \bar D^*/\Xi_b^{\prime} \bar D^*/\Xi_b^{*} \bar D^*$ coupled system with $1(1/2^-)$}\\\hline
$\Lambda$ &$E$  &$r_{\rm RMS}$& P($\Xi_b \bar D^*/\Xi_b^{\prime} \bar D^*/\Xi_b^{*} \bar D^*$)\\
1.50&$-0.48$ &3.50&\textbf{95.63}/2.97/1.40\\
1.53&$-5.31$ &1.08&\textbf{87.78}/8.29/3.93\\
1.55&$-10.91$ &0.76&\textbf{84.06}/10.80/5.14\\\midrule[1.0pt]
\multicolumn{4}{c}{$\Xi_b \bar D^*/\Xi_b^* \bar D/\Xi_b^{\prime} \bar D^*/\Xi_b^{*} \bar D^*$ coupled system with $1(3/2^-)$}\\\hline
$\Lambda$ &$E$  &$r_{\rm RMS}$& P($\Xi_b \bar D^*/\Xi_b^* \bar D/\Xi_b^{\prime} \bar D^*/\Xi_b^{*} \bar D^*$)\\
1.49&$-0.54$ &3.34&\textbf{94.60}/1.12/0.52/3.76\\
1.52&$-5.06$ &1.12&\textbf{86.43}/2.17/1.29/10.11\\
1.54&$-10.30$ &0.79&\textbf{82.65}/2.31/1.63/13.41\\\midrule[1.0pt]
\multicolumn{4}{c}{$\Xi_b^{\prime} \bar D^*/\Xi_b^{*} \bar D^*$ coupled system with $1(3/2^-)$}\\\hline
$\Lambda$ &$E$  &$r_{\rm RMS}$& P($\Xi_b^{\prime} \bar D^*/\Xi_b^{*} \bar D^*$)\\
1.73&$-0.46$ &3.41&\textbf{95.74}/4.26\\
1.76&$-4.62$ &1.08&\textbf{90.34}/9.66\\
1.78&$-9.04$ &0.78&\textbf{88.50}/11.50\\
\bottomrule[1pt]\bottomrule[1pt]
\end{tabular}
\end{table}

For the isovector $\Xi_b^{(\prime,\,*)} \bar D^{(*)}$ coupled systems, Table~\ref{massspectra5} shows that several isovector $\Xi_b^{(\prime,\,*)} \bar D^{(*)}$ coupled systems can form the loosely bound states, even when no binding is found in the corresponding single-channel analyses. These loosely bound states represent new molecular pentaquark candidates emerging from the channel couplings.

For the $\Xi_b \bar D^*/\Xi_b^{\prime} \bar D^*/\Xi_b^{*} \bar D^*$ coupled system with $I(J^P)=1(1/2^-)$, a loosely bound state appears at $\Lambda = 1.50$ GeV with $E = -0.48$ MeV and $r_{\rm RMS} = 3.50$ fm, dominated by the $\Xi_b \bar D^*$ channel. As the cutoff $\Lambda$ increases to $1.55$ GeV, the binding deepens to $E = -10.91$ MeV and the radius contracts to $0.76$ fm, with the $\Xi_b \bar D^*$  channel remaining the dominant component throughout. This coupled system thus constitutes a possible isovector molecular pentaquark candidate.
For the $\Xi_b \bar D^*/\Xi_b^* \bar D/\Xi_b^{\prime} \bar D^*/\Xi_b^{*} \bar D^*$ coupled system with $I(J^P)=1(3/2^-)$, a loosely bound state emerges at $\Lambda = 1.49$ GeV with $E = -0.54$ MeV and $r_{\rm RMS} = 3.34$ fm, again dominated by the $\Xi_b \bar D^*$ channel. As the cutoff increases to $1.54$ GeV, the binding strengthens to $E = -10.30$ MeV and the radius $r_{\rm RMS}$ reduces to $0.79$ fm, with the $\Xi_b \bar D^*$ channel still dominant at the 82.65\% level. Hence, this coupled system represents another possible isovector molecular pentaquark candidate.
For the $\Xi_b^{\prime} \bar D^*/\Xi_b^{*} \bar D^*$ coupled system with $I(J^P)=1(3/2^-)$, a loosely bound state is observed at $\Lambda = 1.73$ GeV with $E = -0.46$ MeV and $r_{\rm RMS} = 3.41$ fm, primarily composed of the $\Xi_b^{\prime} \bar D^*$ channel. At higher cutoffs of $1.76$ GeV and $1.78$ GeV, the binding energies increase to $E = -4.62$ MeV and $-9.04$ MeV, while the radii $r_{\rm RMS}$ decrease to $1.08$ fm and $0.78$ fm, respectively. The $\Xi_b^{\prime} \bar D^*$ component remains dominant throughout. Although the required cutoff values are somewhat higher than for the previous cases, they still lie within a physically reasonable range. Therefore, this coupled system also qualifies as a possible isovector molecular pentaquark candidate. In summary, after incorporating the coupled-channel effects, we identify three possible isovector molecular pentaquark candidates: the $\Xi_b \bar D^*/\Xi_b^{\prime} \bar D^*/\Xi_b^{*} \bar D^*$ coupled system with $I(J^P)=1(1/2^-)$, the $\Xi_b \bar D^*/\Xi_b^* \bar D/\Xi_b^{\prime} \bar D^*/\Xi_b^{*} \bar D^*$ coupled system with $I(J^P)=1(3/2^-)$, and the $\Xi_b^{\prime} \bar D^*/\Xi_b^{*} \bar D^*$ coupled system with $I(J^P)=1(3/2^-)$.

Within the OBE model, the effective potentials receive contributions from the $\sigma$, $\pi$, $\eta$, $\rho$, and $\omega$ exchanges, and the relative signs and strengths of these contributions depend on the total isospin of the discussed system. For the $\sigma$, $\eta$, and $\omega$ exchanges, the potentials for the isoscalar and isovector pentaquark systems are identical. In contrast, for the $\pi$ and $\rho$ exchanges, the interaction is stronger in the isoscalar pentaquark systems, satisfying the relation $V_{I=0}(\pi/\rho) = -3V_{I=1}(\pi/\rho)$. Consequently, the isoscalar pentaquark systems are more prone to forming the loosely bound states than their isovector counterparts, which implies that a smaller cutoff parameter suffices for the former.

In the above discussion, we have systematically investigated the $\Xi_b^{(\prime,*)} \bar D^{(*)}$ systems as the molecular pentaquark candidates. In the single-channel analysis, several isoscalar $\Xi_b^{(\prime,*)} \bar D^{(*)}$ states form the loosely bound states within reasonable cutoff ranges and emerge as the most promising molecular pentaquark candidates comprising five different flavors: the $\Xi_b \bar D$ state with $I(J^P)=0(1/2^-)$, the $\Xi_b^{\prime} \bar D$ state with $I(J^P)=0(1/2^-)$, the $\Xi_b \bar D^{*}$ states with $I(J^P)=0(1/2^-,\,3/2^-)$, the $\Xi_b^{*} \bar D$ state with $I(J^P)=0(3/2^-)$, the $\Xi_b^{\prime} \bar D^{*}$ states with $I(J^P)=0(1/2^-,\,3/2^-)$, and the $\Xi_b^{*} \bar D^{*}$ states with $I(J^P)=0(1/2^-,\,3/2^-,\,5/2^-)$ \footnote{Here, we wish to emphasize that in the study of the hadronic molecular states, the analysis of their stability is typically performed within a specific channel. In the present work, when we examine the $\Xi_b \bar{D}^*$ hadronic molecular state, we focus specifically on the $\Xi_b \bar{D}^*$ channel. This approach is analogous to the widely accepted interpretation of the experimentally observed $P_c(4440)$ and $P_c(4457)$ states as the $\Sigma_c \bar{D}^*$ hadronic molecules, which is also based on the $\Sigma_c \bar{D}^*$ channel rather than the $\Sigma_c \bar{D}$ channel \cite{Chen:2019asm}.}. Notably, the $\Xi_b \bar D^{*}$ bound states with $I(J^P)=0(1/2^-)$ and $0(3/2^-)$ are degenerate due to the absence of the spin-dependent interactions, whereas the $\Xi_b^{\prime} \bar D^{*}$ and $\Xi_b^{*} \bar D^{*}$ multiplets exhibit clear spin splittings. Additionally, the $\Xi_b^{*} \bar D^{*}$ state with $I(J^P)=1(5/2^-)$ can be considered as a possible isovector molecular pentaquark candidate. When the coupled-channel effects are taken into account, the degeneracy between the $\Xi_b \bar D^{*}$ bound states with $I(J^P)=0(1/2^-,\,3/2^-)$ is lifted, and the properties of the isoscalar $\Xi_b^{(\prime,*)} \bar D^{(*)}$ bound states are modified. Moreover, channel couplings generate new loosely bound states in the isovector $\Xi_b^{(\prime,*)} \bar D^{(*)}$ sector, including the $\Xi_b \bar D^*/\Xi_b^{\prime} \bar D^*/\Xi_b^{*} \bar D^*$ coupled system with $I(J^P)=1(1/2^-)$, the $\Xi_b \bar D^*/\Xi_b^* \bar D/\Xi_b^{\prime} \bar D^*/\Xi_b^{*} \bar D^*$ coupled system with $I(J^P)=1(3/2^-)$, and the $\Xi_b^{\prime} \bar D^*/\Xi_b^{*} \bar D^*$ coupled system with $I(J^P)=1(3/2^-)$. These systems represent the isovector molecular pentaquark candidates emerging from the channel couplings.

\section{$\Xi_c^{(\prime,\,*)} B^{(*)}$ systems}\label{sec3}

Building upon our analysis of the $\Xi_b^{(\prime,*)}\bar D^{(*)}$ systems, we now turn to investigate the bound state properties of the $\Xi_c^{(\prime,*)} B^{(*)}$ systems.
In the heavy quark limit, the heavy quark flavor symmetry implies that the low-energy interactions of a heavy-light hadron are independent of the heavy quark flavor \cite{Isgur:1989vq,Isgur:1990yhj}. Within the framework of the heavy quark flavor symmetry, the interaction potentials governing the $\Xi_c^{(\prime,*)} B^{(*)}$ systems and the $\Xi_b^{(\prime,*)}\bar D^{(*)}$ systems are equivalent up to the different masses of the constituent hadrons. This symmetry provides a reliable foundation for extrapolating predictions across the heavy-flavor sectors \cite{Liu:2024uxn}. In this section, we systematically explore the bound state properties of the $\Xi_c^{(\prime,*)} B^{(*)}$ systems to identify potential hadronic molecular candidates comprising five different flavors.

\renewcommand\tabcolsep{0.15cm}
\renewcommand{\arraystretch}{1.50}
\begin{table}[!htbp]
\caption{Bound state properties for the $\Xi_c^{(\prime,\,*)} B^{(*)}$ systems in the single-channel analysis incorporating the $S$-$D$ wave mixing dynamics.}\label{massspectra6}
\begin{tabular}{ccc|ccc}\toprule[1pt]\toprule[1pt]
$\Lambda\,{\rm (GeV)}$ &$E\,{\rm (MeV)}$  &$r_{\rm RMS}\,{\rm (fm)}$&$\Lambda\,{\rm (GeV)}$ &$E\,{\rm (MeV)}$  &$r_{\rm RMS}\,{\rm (fm)}$\\\midrule[1.0pt]
\multicolumn{3}{c|}{$\Xi_c B[0(\frac{1}{2}^{-})]$}&\multicolumn{3}{c}{$\Xi_c^{\prime} B[0(\frac{1}{2}^{-})]$}\\\hline
$1.17$&$-0.31$&$4.31$&$1.19$&$-0.33$&$4.16$\\
$1.29$&$-3.99$&$1.50$&$1.32$&$-4.34$&$1.43$\\
$1.41$&$-11.06$&$0.99$&$1.44$&$-11.28$&$0.97$\\\midrule[1.0pt]
\multicolumn{3}{c|}{$\Xi_c B^*[0(\frac{1}{2}^{-})]$}&\multicolumn{3}{c}{$\Xi_c B^*[0(\frac{3}{2}^{-})]$}\\\hline
$1.17$&$-0.32$&$4.24$&$1.17$&$-0.32$&$4.24$\\
$1.29$&$-4.05$&$1.49$&$1.29$&$-4.05$&$1.49$\\
$1.41$&$-11.16$&$0.99$&$1.41$&$-11.16$&$0.99$\\\midrule[1.0pt]
\multicolumn{3}{c|}{$\Xi_c^{\prime} B^*[0(\frac{1}{2}^{-})]$}&\multicolumn{3}{c}{$\Xi_c^{\prime} B^*[0(\frac{3}{2}^{-})]$}\\\hline
$0.79$&$-0.63$&$3.12$&$1.21$&$-0.34$&$4.34$\\
$0.85$&$-4.26$&$1.38$&$1.48$&$-4.84$&$1.55$\\
$0.90$&$-11.13$&$0.94$&$1.74$&$-12.36$&$1.11$\\\midrule[1.0pt]
\multicolumn{3}{c|}{$\Xi_c^{*} B[0(\frac{3}{2}^{-})]$}&\multicolumn{3}{c}{$\Xi_c^{*} B^*[0(\frac{1}{2}^{-})]$}\\\hline
$1.18$&$-0.29$&$4.33$&$0.79$&$-5.75$&$1.24$\\
$1.31$&$-4.25$&$1.43$&$0.82$&$-9.28$&$1.03$\\
$1.43$&$-11.24$&$0.97$&$0.84$&$-12.49$&$0.92$\\\midrule[1.0pt]
\multicolumn{3}{c|}{$\Xi_c^{*} B^*[0(\frac{3}{2}^{-})]$}&\multicolumn{3}{c}{$\Xi_c^{*} B^*[0(\frac{5}{2}^{-})]$}\\\hline
$0.88$&$-0.24$&$4.55$&$1.19$&$-0.33$&$4.45$\\
$0.96$&$-4.13$&$1.46$&$1.46$&$-4.63$&$1.63$\\
$1.03$&$-12.36$&$0.94$&$1.72$&$-12.29$&$1.15$\\\midrule[1.0pt]
\multicolumn{3}{c|}{$\Xi_c^{*} B^*[1(\frac{5}{2}^{-})]$}\\\Xcline{1-3}{0.50pt}
$1.59$&$-0.28$&$3.96$\\
$1.62$&$-3.17$&$1.26$\\
$1.64$&$-6.49$&$0.89$\\
\bottomrule[1pt]\bottomrule[1pt]
\end{tabular}
\end{table}

We now discuss the bound state properties of the $\Xi_c^{(\prime,*)} B^{(*)}$ systems obtained from the single-channel analysis incorporating the $S$-$D$ wave mixing dynamics, as summarized in Table~\ref{massspectra6}. For each system, three representative cutoff values are presented, and the probabilities of individual $S$- and $D$-wave components are not listed here.

The $\Xi_c B$ state with $I(J^P)=0(1/2^-)$ forms a loosely bound state at $\Lambda = 1.17$ GeV, with binding energy $E = -0.31$ MeV and root-mean-square radius $r_{\rm RMS} = 4.31$ fm. As the cutoff $\Lambda$ increases to $1.41$ GeV, the binding deepens to $-11.06$ MeV, accompanied by a corresponding contraction of the radius to $0.99$ fm. Thus, the $\Xi_c B$ state with $I(J^P)=0(1/2^-)$ can form a loosely bound state within a reasonable cutoff range and emerges as a most promising molecular pentaquark candidate comprising five different flavors.

Similarly, the $\Xi_c^{\prime} B$ state with $I(J^P)=0(1/2^-)$ binds at $\Lambda = 1.19$ GeV, with $E = -0.33$ MeV and $r_{\rm RMS} = 4.16$ fm. At $\Lambda = 1.44$ GeV, the binding increases to $-11.28$ MeV and the radius reduces to $0.97$ fm. This state also qualifies as a most promising molecular pentaquark candidate comprising five different flavors.

For the $\Xi_c B^{*}$ bound states with $I(J^P)=0(1/2^-)$ and $0(3/2^-)$, the binding properties are identical due to the absence of the spin-dependent interactions in this system. At $\Lambda = 1.17$ GeV, both bound states exhibit $E = -0.32$ MeV and $r_{\rm RMS} = 4.24$ fm. As the cutoffs increase to $1.29$ GeV and $1.41$ GeV, the binding energies become $-4.05$ MeV and $-11.16$ MeV, with the radii $1.49$ fm and $0.99$ fm, respectively. These states are themselves most promising molecular pentaquark candidates comprising five different flavors and display binding characteristics similar to the $\Xi_c B$ bound state with $I(J^P)=0(1/2^-)$, reflecting nearly identical interactions apart from a slight increase in the reduced mass.

The $\Xi_c^{*} B$ state with $I(J^P)=0(3/2^-)$ binds at $\Lambda = 1.18$ GeV, with $E = -0.29$ MeV and $r_{\rm RMS} = 4.33$ fm. At larger cutoffs of $1.31$ GeV and $1.43$ GeV, the binding energies deepen to $-4.25$ MeV and $-11.24$ MeV, with the radii $1.43$ fm and $0.97$ fm. This state likewise emerges as a most promising molecular pentaquark candidate comprising five different flavors.

For the $\Xi_c^{\prime} B^{*}$ system, we observe a clear manifestation of the spin-dependent forces. The $\Xi_c^{\prime} B^{*}$ state with $I(J^P)=0(1/2^-)$ binds at a remarkably low cutoff $\Lambda = 0.79$ GeV, with $E = -0.63$ MeV and $r_{\rm RMS} = 3.12$ fm, indicating particularly strong attraction. As the cutoff increases to $0.90$ GeV, the binding deepens to $-11.13$ MeV and the radius shrinks to $0.94$ fm. In contrast, the $\Xi_c^{\prime} B^{*}$ state with $I(J^P)=0(3/2^-)$ requires a significantly larger cutoff $\Lambda = 1.21$ GeV to bind, with $E = -0.34$ MeV and $r_{\rm RMS} = 4.34$ fm. At higher cutoff of $1.74$ GeV, the binding reaches $-12.36$ MeV. This pronounced difference in critical cutoff values directly reflects the influence of the spin-dependent interactions. The $\Xi_c^{\prime} B^{*}$ states with $I(J^P)=0(1/2^-,\,3/2^-)$ form the loosely bound states within reasonable cutoff ranges and represent the most promising molecular pentaquark candidates comprising five different flavors.

The $\Xi_c^{*} B^{*}$ system exhibits an even richer structure, with three distinct total spin-parity quantum numbers: $1/2^-$, $3/2^-$, and $5/2^-$. The spin-dependent interactions lead to markedly different binding behaviors:
\begin{itemize}
    \item The $\Xi_c^{*} B^{*}$ bound state with $I(J^P)=0(1/2^-)$ already exists at $\Lambda = 0.79$ GeV. As the cutoff $\Lambda$ increases to $0.84$ GeV, the binding further deepens to $-12.49$ MeV, with the radius $0.92$ fm.
    \item The $\Xi_c^{*} B^{*}$ bound state with $I(J^P)=0(3/2^-)$ appears at $\Lambda = 0.88$ GeV with $E = -0.24$ MeV and $r_{\rm RMS} = 4.55$ fm. At $\Lambda = 1.03$ GeV, the binding becomes $-12.36$ MeV, with the radius $0.94$ fm.
    \item The $\Xi_c^{*} B^{*}$ state with $I(J^P)=0(5/2^-)$ requires a larger cutoff $\Lambda = 1.19$ GeV to bind, with $E = -0.33$ MeV and $r_{\rm RMS} = 4.45$ fm. At $\Lambda =1.72$ GeV, the binding deepens to $-12.29$ MeV, with the radius $1.15$ fm. This bound state requires the largest cutoff among the three, reflecting the spin dependence of the interaction potentials.
\end{itemize}
All three isoscalar $\Xi_c^{*} B^{*}$ states form the loosely bound states within reasonable cutoff ranges and therefore emerge as the most promising molecular pentaquark candidates comprising five different flavors.
For the isovector $\Xi_c^{*} B^{*}$ system, the $\Xi_c^{*} B^{*}$ state with $I(J^P)=1(5/2^-)$  binds at $\Lambda = 1.59$ GeV with $E = -0.28$ MeV and $r_{\rm RMS} = 3.96$ fm. Although the required cutoff is somewhat larger than for its isoscalar partners, it still lies within a reasonable range. The $\Xi_c^{*} B^{*}$ state with $I(J^P)=1(5/2^-)$ can therefore be considered as a possible isovector molecular pentaquark candidate.

In summary, the single-channel analysis reveals that several isoscalar $\Xi_c^{(\prime,*)} B^{(*)}$ states form the loosely bound states within reasonable cutoff ranges, establishing them as the most promising molecular pentaquark candidates comprising five different flavors: the $\Xi_c B$ state with $I(J^P)=0(1/2^-)$, the $\Xi_c^{\prime} B$ state with $I(J^P)=0(1/2^-)$, the $\Xi_c B^{*}$ states with $I(J^P)=0(1/2^-,\,3/2^-)$, the $\Xi_c^{*} B$ state with $I(J^P)=0(3/2^-)$, the $\Xi_c^{\prime} B^{*}$ states with $I(J^P)=0(1/2^-,\,3/2^-)$, and the $\Xi_c^{*} B^{*}$ states with $I(J^P)=0(1/2^-,\,3/2^-,\,5/2^-)$. In addition, the $\Xi_c^{*} B^{*}$ state with $I(J^P)=1(5/2^-)$ can be considered as a possible isovector molecular pentaquark candidate. A key observation is the degeneracy of the $\Xi_c B^*$ bound states with different total angular momenta in the single-channel analysis, which arises from the absence of the spin-dependent interactions in this channel. In contrast, the $\Xi_c^{\prime} B^*$ and $\Xi_c^* B^*$ bound states exhibit clear spin splittings, providing direct evidence of the role played by the spin-dependent interactions in these systems.

In the following, we extend our investigation to discuss the bound state properties for the $\Xi_c^{(\prime,\,*)} B^{(*)}$ systems obtained from the coupled-channel calculations, which are performed for both the isoscalar and isovector sectors. For the isoscalar $\Xi_c^{(\prime,\,*)} B^{(*)}$ coupled systems, the coupled-channel effects modify the single-channel results and, most notably, remove the degeneracy observed in the single-channel $\Xi_c B^*$ bound states with $I(J^P)=0(1/2^-)$ and $0(3/2^-)$. For the isovector $\Xi_c^{(\prime,\,*)} B^{(*)}$ coupled systems, the coupled-channel dynamics generate new loosely bound states that were absent in the single-channel analysis, thereby expanding the spectrum of the possible molecular pentaquark candidates.

\renewcommand\tabcolsep{0.16cm}
\renewcommand{\arraystretch}{1.50}
\begin{table}[!htbp]
\caption{Bound state properties for the isoscalar $\Xi_c^{(\prime,\,*)} B^{(*)}$ systems in the coupled-channel calculations. The dominant channel for each bound state is denoted in bold.}\label{massspectra7}
\begin{tabular}{cccc}\toprule[1pt]\toprule[1pt]
\multicolumn{4}{c}{$\Xi_c B/\Xi_c B^*/\Xi_c^{\prime} B/\Xi_c^{\prime} B^*/\Xi_c^{*} B^*$ coupled system with $0(1/2^-)$}\\\hline
$\Lambda\,{\rm (GeV)}$ &$E\,{\rm (MeV)}$  &$r_{\rm RMS}\,{\rm (fm)}$& P($\Xi_c B/\Xi_c B^*/\Xi_c^{\prime} B/\Xi_c^{\prime} B^*/\Xi_c^{*} B^*$)\\
0.88&$-0.14$ &5.07&\textbf{96.89}/0.25/0.11/1.81/0.94\\
0.91&$-3.65$ &1.32&\textbf{78.86}/3.66/1.63/12.62/3.23\\
0.93&$-9.42$ &0.80&\textbf{57.56}/10.50/4.73/24.45/2.76\\\midrule[1.0pt]
\multicolumn{4}{c}{$\Xi_c B^*/\Xi_c^{\prime} B/\Xi_c^{\prime} B^*/\Xi_c^{*} B^*$ coupled system with $0(1/2^-)$}\\\hline
$\Lambda\,{\rm (GeV)}$ &$E\,{\rm (MeV)}$  &$r_{\rm RMS}\,{\rm (fm)}$& P($\Xi_c B^*/\Xi_c^{\prime} B/\Xi_c^{\prime} B^*/\Xi_c^{*} B^*$)\\
0.80&$-0.17$ &4.81&\textbf{94.68}/2.56/2.30/0.46\\
0.83&$-3.71$ &1.31&\textbf{78.05}/10.17/9.85/1.93\\
0.86&$-11.41$ &0.78&\textbf{65.24}/15.33/16.29/3.14\\\midrule[1.0pt]
\multicolumn{4}{c}{$\Xi_c B^*/\Xi_c^{\prime} B^*/\Xi_c^* B/\Xi_c^{*} B^*$ coupled system with $0(3/2^-)$}\\\hline
$\Lambda\,{\rm (GeV)}$ &$E\,{\rm (MeV)}$  &$r_{\rm RMS}\,{\rm (fm)}$& P($\Xi_c B^*/\Xi_c^{\prime} B^*/\Xi_c^* B/\Xi_c^{*} B^*$)\\
0.87&$-0.56$ &3.28&\textbf{95.20}/0.38/1.71/2.71\\
0.90&$-4.76$ &1.22&\textbf{85.38}/1.17/4.91/8.53\\
0.92&$-9.94$ &0.87&\textbf{78.89}/1.72/6.78/12.61\\\midrule[1.0pt]
\multicolumn{4}{c}{$\Xi_c^{\prime} B/\Xi_c^{\prime} B^*/\Xi_c^{*} B^*$ coupled system with $0(1/2^-)$}\\\hline
$\Lambda\,{\rm (GeV)}$ &$E\,{\rm (MeV)}$  &$r_{\rm RMS}\,{\rm (fm)}$& P($\Xi_c^{\prime} B/\Xi_c^{\prime} B^*/\Xi_c^{*} B^*$)\\
0.85&$-0.15$ &4.92&\textbf{93.47}/6.31/0.22\\
0.88&$-2.93$ &1.41&\textbf{74.09}/25.42/0.49\\
0.91&$-8.71$ &0.85&\textbf{58.57}/41.14/0.29\\\midrule[1.0pt]
\multicolumn{4}{c}{$\Xi_c^{\prime} B^*/\Xi_c^{*} B^*$ coupled system with $0(1/2^-)$}\\\hline
$\Lambda\,{\rm (GeV)}$ &$E\,{\rm (MeV)}$  &$r_{\rm RMS}\,{\rm (fm)}$& P($\Xi_c^{\prime} B^*/\Xi_c^{*} B^*$)\\
0.79&$-0.91$ &2.62&\textbf{98.89}/1.11\\
0.84&$-5.05$ &1.24&\textbf{96.13}/3.87\\
0.88&$-12.43$ &0.86&\textbf{91.55}/8.45\\\midrule[1.0pt]
\multicolumn{4}{c}{$\Xi_c^* B/\Xi_c^{*} B^*$ coupled system with $0(3/2^-)$}\\\hline
$\Lambda\,{\rm (GeV)}$ &$E\,{\rm (MeV)}$  &$r_{\rm RMS}\,{\rm (fm)}$& P($\Xi_c^* B/\Xi_c^{*} B^*$)\\
0.89&$-0.41$ &3.62&\textbf{92.45}/7.55\\
0.93&$-3.90$ &1.28&\textbf{77.92}/22.08\\
0.97&$-10.19$ &0.84&\textbf{67.10}/32.90\\
\bottomrule[1pt]\bottomrule[1pt]
\end{tabular}
\end{table}

Table~\ref{massspectra7} summarizes the bound state solutions for the isoscalar $\Xi_c^{(\prime,\,*)} B^{(*)}$ systems obtained from the coupled-channel calculations. The coupled-channel effects modify the single-channel results and introduce the channel mixing. The most notable observation is the lifting of the degeneracy present in the single-channel $\Xi_c B^*$ bound states with $I(J^P)=0(1/2^-)$ and $0(3/2^-)$. Specifically, the $\Xi_c B^*/\Xi_c^{\prime} B/\Xi_c^{\prime} B^*/\Xi_c^{*} B^*$ coupled system with $I(J^P)=0(1/2^-)$ and the $\Xi_c B^*/\Xi_c^{\prime} B^*/\Xi_c^* B/\Xi_c^{*} B^*$ coupled system with $I(J^P)=0(3/2^-)$ exhibit different bound state properties at comparable cutoffs, clearly breaking the degeneracy.

\renewcommand\tabcolsep{0.25cm}
\renewcommand{\arraystretch}{1.50}
\begin{table}[!htbp]
\caption{Bound state properties for the isovector $\Xi_c^{(\prime,\,*)} B^{(*)}$ systems in the coupled-channel calculations. The dominant channel for each bound state is denoted in bold.}\label{massspectra8}
\begin{tabular}{cccc}\toprule[1pt]\toprule[1pt]
\multicolumn{4}{c}{$\Xi_c B^*/\Xi_c^{\prime} B/\Xi_c^{\prime} B^*/\Xi_c^{*} B^*$ coupled system with $1(1/2^-)$}\\\hline
$\Lambda\,{\rm (GeV)}$ &$E\,{\rm (MeV)}$  &$r_{\rm RMS}\,{\rm (fm)}$& P($\Xi_c B^*/\Xi_c^{\prime} B/\Xi_c^{\prime} B^*/\Xi_c^{*} B^*$)\\
1.41&$-0.36$ &3.76&\textbf{95.53}/0.51/2.84/1.12\\
1.44&$-4.64$ &1.09&\textbf{86.35}/1.48/8.70/3.47\\
1.46&$-9.82$ &0.76&\textbf{81.92}/1.90/11.53/4.64\\\midrule[1.0pt]
\multicolumn{4}{c}{$\Xi_c B^*/\Xi_c^{\prime} B^*/\Xi_c^* B/\Xi_c^{*} B^*$ coupled system with $1(3/2^-)$}\\\hline
$\Lambda\,{\rm (GeV)}$ &$E\,{\rm (MeV)}$  &$r_{\rm RMS}\,{\rm (fm)}$& P($\Xi_c B^*/\Xi_c^{\prime} B^*/\Xi_c^* B/\Xi_c^{*} B^*$)\\
1.44&$-0.32$ &3.92&\textbf{96.38}/0.74/0.14/2.75\\
1.47&$-4.63$ &1.10&\textbf{88.37}/2.27/0.41/8.95\\
1.49&$-9.89$ &0.75&\textbf{84.49}/2.92/0.52/12.08\\\midrule[1.0pt]
\multicolumn{4}{c}{$\Xi_c^{\prime} B^*/\Xi_c^* B/\Xi_c^{*} B^*$ coupled system with $1(3/2^-)$}\\\hline
$\Lambda\,{\rm (GeV)}$ &$E\,{\rm (MeV)}$  &$r_{\rm RMS}\,{\rm (fm)}$& P($\Xi_c^{\prime} B^*/\Xi_c^* B/\Xi_c^{*} B^*$)\\
1.53&$-0.78$ &2.49&\textbf{93.06}/2.98/3.96\\
1.55&$-5.03$ &0.97&\textbf{85.70}/6.40/7.90\\
1.56&$-8.14$ &0.76&\textbf{83.08}/7.68/9.24\\
\bottomrule[1pt]\bottomrule[1pt]
\end{tabular}
\end{table}

Table~\ref{massspectra8} presents the bound state solutions for the isovector $\Xi_c^{(\prime,\,*)} B^{(*)}$ systems obtained from the coupled-channel calculations. Although the required cutoff values are generally larger than those in the $I=0$ sector, several $\Xi_c^{(\prime,\,*)} B^{(*)}$ coupled systems exhibit the loosely bound states within the reasonable range. Specifically, the $\Xi_c B^*/\Xi_c^{\prime} B/\Xi_c^{\prime} B^*/\Xi_c^{*} B^*$ coupled system with $I(J^P)=1(1/2^-)$ binds at $\Lambda = 1.41$ GeV with $E = -0.36$ MeV, dominated by the $\Xi_c B^*$ channel. The $\Xi_c B^*/\Xi_c^{\prime} B^*/\Xi_c^* B/\Xi_c^{*} B^*$ coupled system with $I(J^P)=1(3/2^-)$ binds at $\Lambda = 1.44$ GeV with $E = -0.32$ MeV, also dominated by the $\Xi_c B^*$ channel. The $\Xi_c^{\prime} B^*/\Xi_c^* B/\Xi_c^{*} B^*$ coupled system with $I(J^P)=1(3/2^-)$ requires a slightly larger cutoff $\Lambda = 1.53$ GeV, binding with $E = -0.78$ MeV and dominated by the $\Xi_c^{\prime} B^*$ channel. These results establish the $\Xi_c B^*/\Xi_c^{\prime} B/\Xi_c^{\prime} B^*/\Xi_c^{*} B^*$ coupled system with $I(J^P)=1(1/2^-)$, the $\Xi_c B^*/\Xi_c^{\prime} B^*/\Xi_c^* B/\Xi_c^{*} B^*$ coupled system with $I(J^P)=1(3/2^-)$, and the $\Xi_c^{\prime} B^*/\Xi_c^* B/\Xi_c^{*} B^*$ coupled system with $I(J^P)=1(3/2^-)$ as the  possible isovector molecular pentaquark candidates.

\begin{figure}[htbp]
  \centering
  \includegraphics[width=8.6cm]{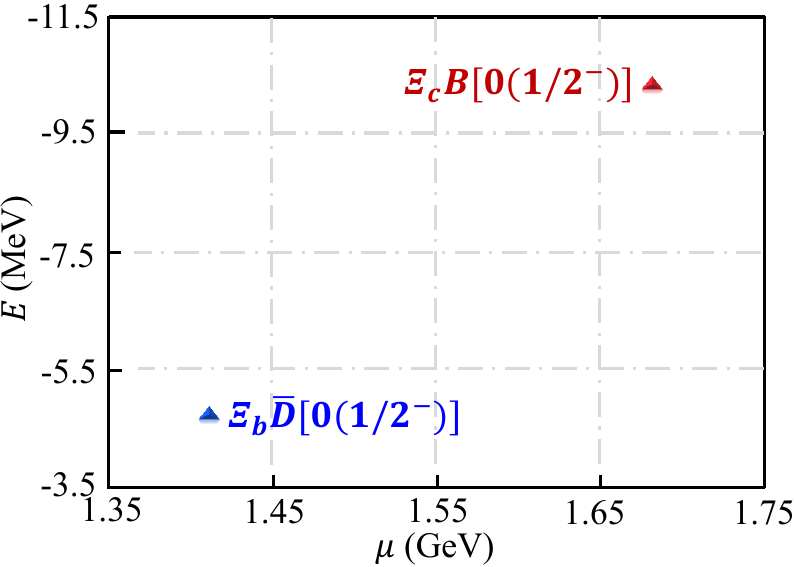}
  \caption{Relationship between the reduced mass and the binding energy under the same interaction strength, taking the $\Xi_b\bar D$ and $\Xi_c B$ bound states with $I(J^P)=0(1/2^-)$ as examples.}\label{reducedmass}
\end{figure}

Finally, we discuss the relationship between the bound state properties of the $\Xi_b^{(\prime,*)}\bar D^{(*)}$ and $\Xi_c^{(\prime,*)} B^{(*)}$ systems. As a direct consequence of the heavy quark flavor symmetry, the $\Xi_b^{(\prime,*)}\bar D^{(*)}$ and $\Xi_c^{(\prime,*)} B^{(*)}$ systems exhibit qualitatively analogous binding behaviors. However, from a quantitative perspective, the binding energies of the symmetry-related systems diverge significantly due to the disparity in their reduced masses. For example, the reduced mass of the $\Xi_b\bar D$ system, approximately $\mu_{\Xi_b\bar D} \approx 1.41\,\text{GeV}$, is notably smaller than that of the $\Xi_c B$ system, with $\mu_{\Xi_c B} \approx 1.68\,\text{GeV}$. Under the same interaction strength, this reduced-mass effect leads to considerably weaker binding in the $\Xi_b\bar D$ state compared to the $\Xi_c B$ state with the same quantum numbers $I(J^P)=0(1/2^-)$, as illustrated in Fig.~\ref{reducedmass}. Concretely, at a cutoff scale $\Lambda = 1.40\,\text{GeV}$, the binding energy of the $\Xi_b\bar D$ bound state with quantum number $I(J^P) = 0(1/2^-)$ is approximately $4.86\,\text{MeV}$, whereas the $\Xi_c B$ state with identical quantum number binds with an energy of around $10.37\,\text{MeV}$. This trend persists across all other heavy flavor partners considered in this work, underscoring the role of the kinematic factors in modulating the dynamics of the heavy-flavor molecular candidates \cite{Li:2014gra,Karliner:2015ina}.

\section{Summary and outlook}\label{sec4}

The discovery of the hidden-charm pentaquarks has established the existence of the hadronic molecules in the heavy-flavor sector, while the observation of the $X_0(2900)$ and $X_1(2900)$ has provided important evidence for  exotic hadrons comprising four different flavors. Together, these discoveries motivate the search for even more exotic hadron configurations. In this work, we performed a systematic investigation of the molecular pentaquark candidates composed of five different flavors, focusing on the $\Xi_b^{(\prime,\,*)} \bar D^{(*)}$ and $\Xi_c^{(\prime,\,*)} B^{(*)}$ systems. Such configurations, containing a bottom quark, a charm quark, a strange quark, an up quark, and a down quark, represent genuinely exotic hadrons. Using the OBE model, we derive the effective interactions between the bottom-strange baryon $\Xi_b^{(\prime,\,*)}$ and an anti-charmed meson $\bar D^{(*)}$, incorporating both the $S$-$D$ wave mixing and coupled-channel dynamics. By solving the coupled-channel Schr\"odinger equation, we explore the formation of the loosely bound states in the  $\Xi_b^{(\prime,\,*)} \bar D^{(*)}$ systems and identify the molecular pentaquark candidates.

Several isoscalar $\Xi_b^{(\prime,\,*)} \bar D^{(*)}$ bound states emerge as the most promising molecular pentaquark candidates comprising five different flavors. These include the $\Xi_b \bar D$ state with $I(J^P)=0(1/2^-)$, the $\Xi_b^{\prime} \bar D$ state with $I(J^P)=0(1/2^-)$, the $\Xi_b \bar D^{*}$ states with $I(J^P)=0(1/2^-,\,3/2^-)$, the $\Xi_b^{*} \bar D$ state with $I(J^P)=0(3/2^-)$, the $\Xi_b^{\prime} \bar D^{*}$ states with $I(J^P)=0(1/2^-,\,3/2^-)$, and the $\Xi_b^{*} \bar D^{*}$ states with $I(J^P)=0(1/2^-,\,3/2^-,\,5/2^-)$. In addition, the $\Xi_b^{*} \bar D^{*}$ state with $I(J^P)=1(5/2^-)$ can be considered a possible isovector molecular pentaquark candidate. In the single-channel analysis, the $\Xi_b \bar D^{*}$ bound states with different total angular momenta are degenerate due to the absence of spin-dependent interactions, whereas the $\Xi_b^{\prime} \bar D^{*}$ and $\Xi_b^{*} \bar D^{*}$ multiplets exhibit clear spin splittings. The coupled-channel effects lift this degeneracy and generate new loosely bound states in the isovector $\Xi_b^{(\prime,\,*)} \bar D^{(*)}$ sector, specifically the $\Xi_b \bar D^*/\Xi_b^{\prime} \bar D^*/\Xi_b^{*} \bar D^*$ coupled system with $I(J^P)=1(1/2^-)$, the $\Xi_b \bar D^*/\Xi_b^* \bar D/\Xi_b^{\prime} \bar D^*/\Xi_b^{*} \bar D^*$ coupled system with $I(J^P)=1(3/2^-)$, and the $\Xi_b^{\prime} \bar D^*/\Xi_b^{*} \bar D^*$ coupled system with $I(J^P)=1(3/2^-)$.

By invoking the heavy quark flavor symmetry, we extend our analysis to the $\Xi_c^{(\prime,\,*)} B^{(*)}$ systems. The pattern of the bound state properties mirrors that of their $\Xi_b^{(\prime,\,*)} \bar D^{(*)}$ partners. The single-channel analysis identifies the $\Xi_c B$ state with $I(J^P)=0(1/2^-)$, the $\Xi_c^{\prime} B$ state with $I(J^P)=0(1/2^-)$, the $\Xi_c B^{*}$ states with $I(J^P)=0(1/2^-,\,3/2^-)$, the $\Xi_c^{*} B$ state with $I(J^P)=0(3/2^-)$, the $\Xi_c^{\prime} B^{*}$ states with $I(J^P)=0(1/2^-,\,3/2^-)$, and the $\Xi_c^{*} B^{*}$ states with $I(J^P)=0(1/2^-,\,3/2^-,\,5/2^-)$ as the most promising molecular pentaquark candidates composed of five different flavors, with the $\Xi_c^{*} B^{*}$ state with $I(J^P)=1(5/2^-)$ representing a possible isovector molecular pentaquark candidate. In addition, the coupled-channel effects again produce new loosely bound states, including the $\Xi_c B^*/\Xi_c^{\prime} B/\Xi_c^{\prime} B^*/\Xi_c^{*} B^*$ coupled system with $I(J^P)=1(1/2^-)$, the $\Xi_c B^*/\Xi_c^{\prime} B^*/\Xi_c^* B/\Xi_c^{*} B^*$ coupled system with $I(J^P)=1(3/2^-)$, and the $\Xi_c^{\prime} B^*/\Xi_c^* B/\Xi_c^{*} B^*$ coupled system with $I(J^P)=1(3/2^-)$.

Notably, the $\Xi_b \bar D$ and $\Xi_c B$ molecular states with $I(J^P)=0(1/2^-)$ can be regarded as the bottom partner of the observed $P_{cs}(4338)$ within the $\Xi_c \bar D$ molecular picture. Similarly, the $\Xi_b \bar D^{*}$ and $\Xi_c B^{*}$ molecular states serve as the heavy-quark partners of the observed $P_{cs}(4459)$ in the $\Xi_c \bar D^{*}$ molecular framework. Further investigation of these bottom-sector partners may provide valuable insights into the inner structures of the $P_{cs}(4338)$ and $P_{cs}(4459)$, and deepen our understanding of the hidden-charm molecular pentaquarks through the lens of the heavy-quark flavor symmetry.

In summary, our comprehensive analysis predicts a rich spectrum of molecular pentaquark candidates composed of five different quark flavors. We strongly encourage experimental searches for the most promising candidates identified in this work. Analogous to the discovery of the $P_c$ and $P_{cs}$ states by LHCb \cite{Aaij:2015tga,Aaij:2019vzc,LHCb:2020jpq,LHCb:2022ogu}, these molecular pentaquark candidates could be observed in the invariant mass spectra of the final states containing a $B_c$ meson and a $\Lambda$ baryon. The high luminosity of the proton-proton collisions at LHCb makes it particularly well suited for their direct production. In parallel, inspired by the evidence for the $P_{cs}(4459)$ pentaquark observed with a local significance of $3.3\sigma$ in the inclusive decays of the $\Upsilon(1S,2S)$ at Belle~\cite{Belle:2025pey}, these molecular pentaquark candidates can also be investigated in the decays of the $\Upsilon$ resonances, offering a complementary production mechanism. The distinctive five-flavor quark configuration of the predicted states provides a highly distinguishable ``flavor tag'' for future experimental searches, significantly enhancing their detectability.

\section*{Acknowledgement}

This work is also supported by the Natural Science Foundation of Gansu Province (No. 26RCKA012, No. 25JRRA799), the National Natural Science Foundation of China under Grants No. 12335001, No. 12405097 and No. 12247101, the ‘111 Center’ under Grant No. B20063, the fundamental Research Funds for the Central Universities (lzujbky-2023-stlt01), Lanzhou City High-Level Talent Funding, and the Talent Scientific Fund of Lanzhou University.


\begin{thebibliography}{99}

%\cite{GellMann:1964nj}
\bibitem{GellMann:1964nj}
  M.~Gell-Mann,
  A schematic model of baryons and mesons,
  \href{https://www.sciencedirect.com/science/article/abs/pii/S0031916364920013?via\%3Dihub}{Phys.\ Lett.\  {\bf 8}, 214 (1964)}.
  %%doi:10.1016/S0031-9163(64)92001-3
  %%CITATION = doi:10.1016/S0031-9163(64)92001-3;%%
  %3407 citations counted in INSPIRE as of 04 Mar 2021

%\cite{Zweig:1981pd}
\bibitem{Zweig:1981pd}
G.~Zweig,
An SU(3) model for strong interaction symmetry and its breaking. Version 1,
\href{http://cds.cern.ch/record/352337}{CERN-TH-401.}
%618 citations counted in INSPIRE as of 30 Aug 2020

%\cite{Liu:2013waa}
\bibitem{Liu:2013waa}
      X.~Liu,
      An overview of $XYZ$ new particles,
      \href{http://dx.doi.org/10.1007/s11434-014-0407-2}{Chin.\ Sci.\ Bull.\  {\bf 59}, 3815 (2014)}.
      %doi:10.1007/s11434-014-0407-2
      %\href{https://arxiv.org/abs/1312.7408}{[arXiv:1312.7408 [hep-ph]]}.
      %%CITATION = doi:10.1007/s11434-014-0407-2;%%
      %67 citations counted in INSPIRE as of 19 Jul 2017

%\cite{Chen:2016qju}
\bibitem{Chen:2016qju}
  H.~X.~Chen, W.~Chen, X.~Liu, and S.~L.~Zhu,
  The hidden-charm pentaquark and tetraquark states,
  \href{http://linkinghub.elsevier.com/retrieve/pii/S037015731630103X}{Phys.\ Rep.\  {\bf 639}, 1 (2016)}.
 % doi:10.1016/j.physrep.2016.05.004
 %\href{http://arxiv.org/abs/arXiv:1601.02092}{[arXiv:1601.02092 [hep-ph]]}.
  %%CITATION = doi:10.1016/j.physrep.2016.05.004;%%
  %175 citations counted in INSPIRE as of 19 Jul 2017

%\cite{Hosaka:2016pey}
\bibitem{Hosaka:2016pey}
  A.~Hosaka, T.~Iijima, K.~Miyabayashi, Y.~Sakai, and S.~Yasui,
  Exotic hadrons with heavy flavors: $X$, $Y$, $Z$, and related states,
  \href{http://dx.doi.org/10.1093/ptep/ptw045}{Prog. Theor. Exp. Phys. {\bf 2016}, 062C01 (2016)}.
 % doi:10.1093/ptep/ptw045
 %\href{https://arxiv.org/abs/1603.09229}{[arXiv:1603.09229 [hep-ph]]}.
  %%CITATION = doi:10.1093/ptep/ptw045;%%
  %29 citations counted in INSPIRE as of 25 Jul 2017

\bibitem{Richard:2016eis}
J.~M.~Richard,
Exotic hadrons: review and perspectives,
\href{https://link.springer.com/article/10.1007/s00601-016-1159-0}{Few Body Syst. \textbf{57}, 1185-1212 (2016)}.
%doi:10.1007/s00601-016-1159-0
%[arXiv:1606.08593 [hep-ph]].
%76 citations counted in INSPIRE as of 08 Jun 2021

%\cite{Lebed:2016hpi}
\bibitem{Lebed:2016hpi}
R.~F.~Lebed, R.~E.~Mitchell, and E.~S.~Swanson,
Heavy-Quark QCD Exotica,
\href{https://www.sciencedirect.com/science/article/pii/S0146641016300734?via\%3Dihub}{Prog. Part. Nucl. Phys. \textbf{93}, 143-194 (2017)}.
%doi:10.1016/j.ppnp.2016.11.003
%[arXiv:1610.04528 [hep-ph]].
%435 citations counted in INSPIRE as of 07 Jul 2022

%\cite{Olsen:2017bmm}
\bibitem{Olsen:2017bmm}
  S.~L.~Olsen, T.~Skwarnicki, and D.~Zieminska,
  Nonstandard heavy mesons and baryons: Experimental evidence,
  \href{https://journals.aps.org/rmp/abstract/10.1103/RevModPhys.90.015003}{Rev.\ Mod.\ Phys.\  {\bf 90}, 015003 (2018)}.
  %%doi:10.1103/RevModPhys.90.015003
  %% [arXiv:1708.04012 [hep-ph]].
  %%CITATION = doi:10.1103/RevModPhys.90.015003;%%
  %196 citations counted in INSPIRE as of 17 Dec 2019

%\cite{Guo:2017jvc}
\bibitem{Guo:2017jvc}
  F.~K.~Guo, C.~Hanhart, U.~G.~Mei$\ss$ner, Q.~Wang, Q.~Zhao, and B.~S.~Zou,
  Hadronic molecules,
  \href{https://journals.aps.org/rmp/abstract/10.1103/RevModPhys.90.015004}{Rev.\ Mod.\ Phys.\  {\bf 90}, 015004 (2018)}.
  %%doi:10.1103/RevModPhys.90.015004
  %%[arXiv:1705.00141 [hep-ph]].
  %%CITATION = doi:10.1103/RevModPhys.90.015004;%%
  %322 citations counted in INSPIRE as of 17 Dec 2019

%\cite{Liu:2019zoy}
\bibitem{Liu:2019zoy}
Y.~R.~Liu, H.~X.~Chen, W.~Chen, X.~Liu, and S.~L.~Zhu,
Pentaquark and tetraquark states,
\href{https://www.sciencedirect.com/science/article/pii/S0146641019300304?via\%3Dihub}{Prog.\ Part.\ Nucl.\ Phys.\  {\bf 107}, 237 (2019)}.
%%doi:10.1016/j.ppnp.2019.04.003
%%[arXiv:1903.11976 [hep-ph]].
%%CITATION = doi:10.1016/j.ppnp.2019.04.003;%%
%30 citations counted in INSPIRE as of 29 Jun 2019

%\cite{Brambilla:2019esw}
\bibitem{Brambilla:2019esw}
N.~Brambilla, S.~Eidelman, C.~Hanhart, A.~Nefediev, C.~P.~Shen, C.~E.~Thomas, A.~Vairo, and C.~Z.~Yuan,
The $XYZ$ states: Experimental and theoretical status and perspectives,
\href{https://www.sciencedirect.com/science/article/pii/S0370157320301915?via\%3Dihub}{Phys. Rep. \textbf{873}, 1 (2020)}.
%doi:10.1016/j.physrep.2020.05.001
%[arXiv:1907.07583 [hep-ex]].
%99 citations counted in INSPIRE as of 24 Aug 2020

%\cite{Meng:2022ozq}
\bibitem{Meng:2022ozq}
L.~Meng, B.~Wang, G.~J.~Wang, and S.~L.~Zhu,
Chiral perturbation theory for heavy hadrons and chiral effective field theory for heavy hadronic molecules,
\href{https://www.sciencedirect.com/science/article/pii/S0370157323001679?via\%3Dihub}{Phys. Rept. \textbf{1019}, 1-149 (2023)}.
%doi:10.1016/j.physrep.2023.04.003
%[arXiv:2204.08716 [hep-ph]].
%63 citations counted in INSPIRE as of 02 Jul 2023

%\cite{Chen:2022asf}
\bibitem{Chen:2022asf}
H.~X.~Chen, W.~Chen, X.~Liu, Y.~R.~Liu, and S.~L.~Zhu,
An updated review of the new hadron states,
\href{https://iopscience.iop.org/article/10.1088/1361-6633/aca3b6}{Rept. Prog. Phys. \textbf{86}, 026201 (2023)}.
%doi:10.1088/1361-6633/aca3b6
%[arXiv:2204.02649 [hep-ph]].
%164 citations counted in INSPIRE as of 02 Jul 2023

%\cite{Liu:2024uxn}
\bibitem{Liu:2024uxn}
M.~Z.~Liu, Y.~W.~Pan, Z.~W.~Liu, T.~W.~Wu, J.~X.~Lu, and L.~S.~Geng,
Three ways to decipher the nature of exotic hadrons: Multiplets, three-body hadronic molecules, and correlation functions,
\href{https://www.sciencedirect.com/science/article/pii/S0370157324004277?via\%3Dihub}{Phys. Rept. \textbf{1108}, 1-108 (2025)}.
%doi:10.1016/j.physrep.2024.12.001
%[arXiv:2404.06399 [hep-ph]].
%114 citations counted in INSPIRE as of 07 Oct 2025

%\cite{Wang:2025sic}
\bibitem{Wang:2025sic}
Z.~G.~Wang,
Review of the QCD sum rules for exotic states,
\href{https://journal.hep.com.cn/fop/EN/10.15302/frontphys.2026.016300}{Front. Phys. (Beijing) \textbf{21}, 016300 (2026)}.
%doi:10.15302/frontphys.2026.016300
%[arXiv:2502.11351 [hep-ph]].
%32 citations counted in INSPIRE as of 07 Oct 2025

%\cite{Wang:2025dur}
\bibitem{Wang:2025dur}
X.~Wang, X.~Liu, and Y.~Gao,
Colloquium: Hadron production in open-charm meson pair at $e^+e^-$ collider,
\href{https://arxiv.org/abs/2502.15117}{arXiv:2502.15117}.
%8 citations counted in INSPIRE as of 24 Feb 2026

%\cite{Bai:2026atm}
\bibitem{Bai:2026atm}
Z.~Y.~Bai, D.~Y.~Chen, Qi-Huang, X.~Liu, S.~Q.~Luo, and J.~Z.~Wang,
Unquenched charmonium and beyond,
\href{https://arxiv.org/abs/2602.19887}{arXiv:2602.19887}.
%0 citations counted in INSPIRE as of 23 Mar 2026

%\cite{Aaij:2015tga}
\bibitem{Aaij:2015tga}
  R.~Aaij {\it et al.} [LHCb],
 Observation of $J/\psi p$ resonances consistent with pentaquark states in $\Lambda_b^0 \to J/\psi K^- p$ decays,
 \href{https://journals.aps.org/prl/abstract/10.1103/PhysRevLett.115.072001}{Phys.\ Rev.\ Lett.\  {\bf 115}, 072001 (2015)}.
  %doi:10.1103/PhysRevLett.115.072001
  %[arXiv:1507.03414 [hep-ex]].
  %%CITATION = doi:10.1103/PhysRevLett.115.072001;%%
  %83 citations counted in INSPIRE as of 24 Nov 2015

%\cite{Aaij:2019vzc}
\bibitem{Aaij:2019vzc}
  R.~Aaij {\it et al.} [LHCb],
  Observation of a narrow pentaquark state, $P_c(4312)^+$, and of two-peak structure of the $P_c(4450)^+$,
 \href{https://journals.aps.org/prl/abstract/10.1103/PhysRevLett.122.222001}{Phys.\ Rev.\ Lett.\  {\bf 122}, 222001 (2019)}.
  %%doi:10.1103/PhysRevLett.122.222001
  %%[arXiv:1904.03947 [hep-ex]].
  %%CITATION = doi:10.1103/PhysRevLett.122.222001;%%
  %41 citations counted in INSPIRE as of 29 Jun 2019

%\cite{Wu:2010jy}
\bibitem{Wu:2010jy}
  J.~J.~Wu, R.~Molina, E.~Oset, and B.~S.~Zou,
  Prediction of narrow $N^*$ and $\Lambda^*$ resonances with hidden charm above 4 GeV,
  \href{https://journals.aps.org/prl/abstract/10.1103/PhysRevLett.105.232001}{Phys.\ Rev.\ Lett.\  {\bf 105}, 232001 (2010)}.
  %%doi:10.1103/PhysRevLett.105.232001
  %%[arXiv:1007.0573 [nucl-th]].
  %%CITATION = doi:10.1103/PhysRevLett.105.232001;%%
  %279 citations counted in INSPIRE as of 09 Nov 2020

%\cite{Wang:2011rga}
\bibitem{Wang:2011rga}
  W.~L.~Wang, F.~Huang, Z.~Y.~Zhang, and B.~S.~Zou,
  $\Sigma_c \bar{D}$ and $\Lambda_c \bar{D}$ states in a chiral quark model,
  \href{https://journals.aps.org/prc/abstract/10.1103/PhysRevC.84.015203}{Phys.\ Rev.\ C {\bf 84}, 015203 (2011)}.
  %doi:10.1103/PhysRevC.84.015203
  %[arXiv:1101.0453 [nucl-th]].
  %%CITATION = doi:10.1103/PhysRevC.84.015203;%%
  %72 citations counted in INSPIRE as of 23 Apr 2019

%\cite{Yang:2011wz}
\bibitem{Yang:2011wz}
  Z.~C.~Yang, Z.~F.~Sun, J.~He, X.~Liu, and S.~L.~Zhu,
  The possible hidden-charm molecular baryons composed of anti-charmed meson and charmed baryon,
  \href{https://iopscience.iop.org/article/10.1088/1674-1137/36/1/002/meta}{Chin.\ Phys.\ C {\bf 36}, 6 (2012)}.
  %doi:10.1088/1674-1137/36/1/002, 10.1088/1674-1137/36/3/006
 % [arXiv:1105.2901 [hep-ph]].
  %%CITATION = doi:10.1088/1674-1137/36/1/002, 10.1088/1674-1137/36/3/006;%%
  %103 citations counted in INSPIRE as of 23 Apr 2019

%\cite{Wu:2012md}
\bibitem{Wu:2012md}
  J.~J.~Wu, T.-S.~H.~Lee, and B.~S.~Zou,
  Nucleon resonances with hidden charm in coupled-channel Models,
  \href{https://journals.aps.org/prc/abstract/10.1103/PhysRevC.85.044002}{Phys.\ Rev.\ C {\bf 85}, 044002 (2012)}.
  %doi:10.1103/PhysRevC.85.044002
  %[arXiv:1202.1036 [nucl-th]].
  %%CITATION = doi:10.1103/PhysRevC.85.044002;%%
  %44 citations counted in INSPIRE as of 23 Apr 2019

%\cite{Li:2014gra}
\bibitem{Li:2014gra}
  X.~Q.~Li and X.~Liu,
  A possible global group structure for exotic states,
  \href{https://link.springer.com/article/10.1140\%2Fepjc\%2Fs10052-014-3198-3}{Eur.\ Phys.\ J.\ C {\bf 74}, 3198 (2014)}.
  %%doi:10.1140/epjc/s10052-014-3198-3
  %%[arXiv:1409.3332 [hep-ph]].
  %%CITATION = doi:10.1140/epjc/s10052-014-3198-3;%%
  %20 citations counted in INSPIRE as of 18 Mar 2020

%\cite{Chen:2015loa}
\bibitem{Chen:2015loa}
  R.~Chen, X.~Liu, X.~Q.~Li, and S.~L.~Zhu,
  Identifying exotic hidden-charm pentaquarks,
  \href{https://journals.aps.org/prl/abstract/10.1103/PhysRevLett.115.132002}{Phys.\ Rev.\ Lett.\  {\bf 115}, 132002 (2015)}.
  %doi:10.1103/PhysRevLett.115.132002
  %[arXiv:1507.03704 [hep-ph]].
  %%CITATION = doi:10.1103/PhysRevLett.115.132002;%%
  %26 citations counted in INSPIRE as of 24 Nov 2015

%\cite{Karliner:2015ina}
\bibitem{Karliner:2015ina}
  M.~Karliner and J.~L.~Rosner,
  New exotic meson and baryon resonances from doubly-heavy hadronic molecules,
  \href{https://journals.aps.org/prl/abstract/10.1103/PhysRevLett.115.122001}{Phys.\ Rev.\ Lett.\  {\bf 115}, 122001 (2015)}.
  %%doi:10.1103/PhysRevLett.115.122001
  %%[arXiv:1506.06386 [hep-ph]].
  %%CITATION = doi:10.1103/PhysRevLett.115.122001;%%
  %159 citations counted in INSPIRE as of 18 Mar 2020

%\cite{LHCb:2020jpq}
\bibitem{LHCb:2020jpq}
R.~Aaij \textit{et al.} [LHCb],
Evidence of a $J/\psi\Lambda$ structure and observation of excited $\Xi^-$ states in the $\Xi^-_b \to J/\psi\Lambda K^-$ decay,
\href{https://www.sciencedirect.com/science/article/pii/S2095927321001717?via\%3Dihub}{Sci. Bull. \textbf{66}, 1278-1287 (2021)}.
%doi:10.1016/j.scib.2021.02.030
%[arXiv:2012.10380 [hep-ex]].
%113 citations counted in INSPIRE as of 28 Oct 2022

%\cite{LHCb:2022ogu}
\bibitem{LHCb:2022ogu}
R.~Aaij \textit{et al.} [LHCb],
Observation of a $J/\psi\Lambda$ resonance consistent with a strange pentaquark candidate in $B^-\to J/\psi\Lambda\bar{p}$ decays,
\href{https://journals.aps.org/prl/abstract/10.1103/PhysRevLett.131.031901}{Phys. Rev. Lett. \textbf{131}, 031901 (2023)}.
%doi:10.1103/PhysRevLett.131.031901
%[arXiv:2210.10346 [hep-ex]].
%39 citations counted in INSPIRE as of 09 Aug 2023

%\cite{Hofmann:2005sw}
\bibitem{Hofmann:2005sw}
  J.~Hofmann and M.~F.~M.~Lutz,
  Coupled-channel study of crypto-exotic baryons with charm,
  \href{https://www.sciencedirect.com/science/article/pii/S0375947405010596?via\%3Dihub}{Nucl.\ Phys.\ A {\bf 763}, 90 (2005)}.
  %%doi:10.1016/j.nuclphysa.2005.08.022
  %%[hep-ph/0507071].
  %%CITATION = doi:10.1016/j.nuclphysa.2005.08.022;%%
  %189 citations counted in INSPIRE as of 09 Nov 2020

%\cite{Wu:2010vk}
\bibitem{Wu:2010vk}
J.~J.~Wu, R.~Molina, E.~Oset, and B.~S.~Zou,
Dynamically generated $N^{*}$ and $\Lambda^*$ resonances in the hidden charm sector around 4.3 GeV,
\href{https://journals.aps.org/prc/abstract/10.1103/PhysRevC.84.015202}{Phys. Rev. C \textbf{84}, 015202 (2011)}.
%doi:10.1103/PhysRevC.84.015202
%[arXiv:1011.2399 [nucl-th]].
%280 citations counted in INSPIRE as of 16 Oct 2025

%\cite{Anisovich:2015zqa}
\bibitem{Anisovich:2015zqa}
  V.~V.~Anisovich, M.~A.~Matveev, J.~Nyiri, A.~V.~Sarantsev, and A.~N.~Semenova,
  Nonstrange and strange pentaquarks with hidden charm,
  \href{https://www.worldscientific.com/doi/abs/10.1142/S0217751X15501900}{Int.\ J.\ Mod.\ Phys.\ A {\bf 30}, 1550190 (2015)}.
  %%doi:10.1142/S0217751X15501900
  %%[arXiv:1509.04898 [hep-ph]].
  %%CITATION = doi:10.1142/S0217751X15501900;%%
  %31 citations counted in INSPIRE as of 09 Nov 2020

\bibitem{Chen:2016ryt}
  R.~Chen, J.~He, and X.~Liu,
  Possible strange hidden-charm pentaquarks from $\Sigma_c^{(*)}\bar{D}_s^*$ and $\Xi^{(',*)}_c\bar{D}^*$ interactions, \href{https://iopscience.iop.org/article/10.1088/1674-1137/41/10/103105}{Chin.\ Phys.\ C {\bf 41}, 103105 (2017)}.
  %%doi:10.1088/1674-1137/41/10/103105
  %%[arXiv:1609.03235 [hep-ph]].
  %%CITATION = doi:10.1088/1674-1137/41/10/103105;%%
  %13 citations counted in INSPIRE as of 09 Nov 2020

\bibitem{Weng:2019ynv}
X.~Z.~Weng, X.~L.~Chen, W.~Z.~Deng, and S.~L.~Zhu,
Hidden-charm pentaquarks and $P_c$ states,
\href{https://journals.aps.org/prd/abstract/10.1103/PhysRevD.100.016014}{Phys. Rev. D \textbf{100}, 016014 (2019)}.
%doi:10.1103/PhysRevD.100.016014
%[arXiv:1904.09891 [hep-ph]].
%65 citations counted in INSPIRE as of 21 Jul 2022

%\cite{Xiao:2019gjd}
\bibitem{Xiao:2019gjd}
  C.~W.~Xiao, J.~Nieves, and E.~Oset,
  Prediction of hidden charm strange molecular baryon states with heavy quark spin symmetry,
\href{https://www.sciencedirect.com/science/article/pii/S0370269319307737?via\%3Dihub}{Phys.\ Lett.\ B {\bf 799}, 135051 (2019)}.
  %%doi:10.1016/j.physletb.2019.135051
  %%[arXiv:1906.09010 [hep-ph]].
  %%CITATION = doi:10.1016/j.physletb.2019.135051;%%
  %14 citations counted in INSPIRE as of 09 Nov 2020

 %\cite{Wang:2019nvm}
\bibitem{Wang:2019nvm}
  B.~Wang, L.~Meng, and S.~L.~Zhu,
  Spectrum of the strange hidden charm molecular pentaquarks in chiral effective field theory,
\href{https://journals.aps.org/prd/abstract/10.1103/PhysRevD.101.034018}{Phys.\ Rev.\ D {\bf 101}, 034018 (2020)}.
  %%doi:10.1103/PhysRevD.101.034018
  %%[arXiv:1912.12592 [hep-ph]].
  %%CITATION = doi:10.1103/PhysRevD.101.034018;%%
  %6 citations counted in INSPIRE as of 09 Nov 2020

\bibitem{Chen:2020uif}
H.~X.~Chen, W.~Chen, X.~Liu, and X.~H.~Liu,
Establishing the first hidden-charm pentaquark with strangeness,
\href{https://link.springer.com/article/10.1140/epjc/s10052-021-09196-4}{Eur. Phys. J. C \textbf{81}, 409 (2021)}.
%doi:10.1140/epjc/s10052-021-09196-4
%[arXiv:2011.01079 [hep-ph]].
%51 citations counted in INSPIRE as of 28 Oct 2022

%\cite{Du:2021bgb}
\bibitem{Du:2021bgb}
M.~L.~Du, Z.~H.~Guo, and J.~A.~Oller,
Insights into the nature of the $P_{cs}(4459)$,
\href{https://journals.aps.org/prd/abstract/10.1103/PhysRevD.104.114034}{Phys. Rev. D \textbf{104}, 114034 (2021)}.
%doi:10.1103/PhysRevD.104.114034
%[arXiv:2109.14237 [hep-ph]].
%8 citations counted in INSPIRE as of 21 Jul 2022

%\cite{Dong:2021juy}
\bibitem{Dong:2021juy}
X.~K.~Dong, F.~K.~Guo, and B.~S.~Zou,
A survey of heavy-antiheavy hadronic molecules,
\href{https://pip.nju.edu.cn/EN/10.13725/j.cnki.pip.2021.02.001}{Progr. Phys. \textbf{41}, 65 (2021)}.
%doi:10.13725/j.cnki.pip.2021.02.001
%[arXiv:2101.01021 [hep-ph]].
%81 citations counted in INSPIRE as of 28 Oct 2022

%\cite{Lu:2021irg}
\bibitem{Lu:2021irg}
J.~X.~Lu, M.~Z.~Liu, R.~X.~Shi, and L.~S.~Geng,
Understanding $P_{cs}(4459)$ as a hadronic molecule in the $\Xi_b^- \to J/\psi \Lambda K^-$ decay,
\href{https://journals.aps.org/prd/abstract/10.1103/PhysRevD.104.034022}{Phys. Rev. D \textbf{104}, 034022 (2021)}.
%doi:10.1103/PhysRevD.104.034022
%[arXiv:2104.10303 [hep-ph]].
%23 citations counted in INSPIRE as of 03 Jul 2023

%\cite{Zou:2021sha}
\bibitem{Zou:2021sha}
B.~S.~Zou,
Building up the spectrum of pentaquark states as hadronic molecules,
\href{https://www.sciencedirect.com/science/article/pii/S2095927321002826?via\%3Dihub}{Sci. Bull. \textbf{66}, 1258 (2021)}.
%doi:10.1016/j.scib.2021.04.023
%[arXiv:2103.15273 [hep-ph]].
%10 citations counted in INSPIRE as of 03 Jul 2023

%\cite{Wang:2021itn}
\bibitem{Wang:2021itn}
Z.~G.~Wang and Q.~Xin,
Analysis of hidden-charm pentaquark molecular states with and without strangeness via the QCD sum rules,
\href{https://iopscience.iop.org/article/10.1088/1674-1137/ac2a1d}{Chin. Phys. C \textbf{45}, 123105 (2021)}.
%doi:10.1088/1674-1137/ac2a1d
%[arXiv:2103.08239 [hep-ph]].
%16 citations counted in INSPIRE as of 03 Jul 2023

%\cite{Peng:2020hql}
\bibitem{Peng:2020hql}
F.~Z.~Peng, M.~J.~Yan, M.~S\'anchez S\'anchez, and M.~P.~Valderrama,
The $P_{cs}(4459)$ pentaquark from a combined effective field theory and phenomenological perspective,
\href{https://link.springer.com/article/10.1140/epjc/s10052-021-09416-x}{Eur. Phys. J. C \textbf{81}, 666 (2021)}.
%doi:10.1140/epjc/s10052-021-09416-x
%[arXiv:2011.01915 [hep-ph]].
%48 citations counted in INSPIRE as of 28 Oct 2022

%\cite{Zhu:2021lhd}
\bibitem{Zhu:2021lhd}
J.~T.~Zhu, L.~Q.~Song, and J.~He,
$P_{cs}(4459)$ and other possible molecular states from $\Xi_{c}^{(*)}\bar{D}^{(*)}$ and $\Xi^{\prime}_c\bar{D}^{(*)}$ interactions,
\href{https://journals.aps.org/prd/abstract/10.1103/PhysRevD.103.074007}{Phys. Rev. D \textbf{103}, 074007 (2021)}.
%doi:10.1103/PhysRevD.103.074007
%[arXiv:2101.12441 [hep-ph]].
%27 citations counted in INSPIRE as of 21 Jul 2022

%\cite{Wang:2020eep}
\bibitem{Wang:2020eep}
Z.~G.~Wang,
Analysis of the $P_{cs}(4459)$ as the hidden-charm pentaquark state with QCD sum rules,
\href{https://www.worldscientific.com/doi/abs/10.1142/S0217751X21500718}{Int. J. Mod. Phys. A \textbf{36}, 2150071 (2021)}.
%doi:10.1142/S0217751X21500718
%[arXiv:2011.05102 [hep-ph]].
%44 citations counted in INSPIRE as of 03 Jul 2023

%\cite{Chen:2020kco}
\bibitem{Chen:2020kco}
R.~Chen,
Can the newly reported $P_{cs}(4459)$ be a strange hidden-charm $\Xi_c\bar D^*$ molecular pentaquark?,
\href{https://journals.aps.org/prd/abstract/10.1103/PhysRevD.103.054007}{Phys. Rev. D \textbf{103}, 054007 (2021)}.
%doi:10.1103/PhysRevD.103.054007
%[arXiv:2011.07214 [hep-ph]].
%31 citations counted in INSPIRE as of 21 Jul 2022

%\cite{Xiao:2021rgp}
\bibitem{Xiao:2021rgp}
C.~W.~Xiao, J.~J.~Wu, and B.~S.~Zou,
Molecular nature of $P_{cs} (4459)$ and its heavy quark spin partners,
\href{https://journals.aps.org/prd/abstract/10.1103/PhysRevD.103.054016}{Phys. Rev. D \textbf{103}, 054016 (2021)}.
%doi:10.1103/PhysRevD.103.054016
%[arXiv:2102.02607 [hep-ph]].
%24 citations counted in INSPIRE as of 21 Jul 2022

%\cite{Chen:2021cfl}
\bibitem{Chen:2021cfl}
K.~Chen, R.~Chen, L.~Meng, B.~Wang, and S.~L.~Zhu,
Systematics of the heavy flavor hadronic molecules,
\href{https://link.springer.com/article/10.1140/epjc/s10052-022-10540-5}{Eur. Phys. J. C \textbf{82}, 581 (2022)}.
%doi:10.1140/epjc/s10052-022-10540-5
%[arXiv:2109.13057 [hep-ph]].
%21 citations counted in INSPIRE as of 21 Jul 2022

%\cite{Chen:2022onm}
\bibitem{Chen:2022onm}
R.~Chen and X.~Liu,
Mass behavior of hidden-charm open-strange pentaquarks inspired by the established $P_c$ molecular states,
\href{https://journals.aps.org/prd/abstract/10.1103/PhysRevD.105.014029}{Phys. Rev. D \textbf{105}, 014029 (2022)}.
%doi:10.1103/PhysRevD.105.014029
%[arXiv:2201.07603 [hep-ph]].
%2 citations counted in INSPIRE as of 21 Jul 2022

%\cite{Chen:2021spf}
\bibitem{Chen:2021spf}
K.~Chen, B.~Wang, and S.~L.~Zhu,
Heavy flavor molecular states with strangeness,
\href{https://journals.aps.org/prd/abstract/10.1103/PhysRevD.105.096004}{Phys. Rev. D \textbf{105}, 096004 (2022)}.
%doi:10.1103/PhysRevD.105.096004
%[arXiv:2112.13203 [hep-ph]].
%8 citations counted in INSPIRE as of 21 Jul 2022

%\cite{Hu:2021nvs}
\bibitem{Hu:2021nvs}
X.~Hu and J.~Ping,
Investigation of hidden-charm pentaquarks with strangeness $S=-1$,
\href{https://link.springer.com/article/10.1140/epjc/s10052-022-10047-z}{Eur. Phys. J. C \textbf{82}, 118 (2022)}.
%doi:10.1140/epjc/s10052-022-10047-z
%[arXiv:2109.09972 [hep-ph]].
%6 citations counted in INSPIRE as of 21 Jul 2022

%\cite{Garcilazo:2022edi}
\bibitem{Garcilazo:2022edi}
H.~Garcilazo and A.~Valcarce,
Hidden-flavor pentaquarks,
\href{https://journals.aps.org/prd/abstract/10.1103/PhysRevD.106.114012}{Phys. Rev. D \textbf{106}, 114012 (2022)}.
%doi:10.1103/PhysRevD.106.114012
%[arXiv:2211.13949 [hep-ph]].
%2 citations counted in INSPIRE as of 03 Jul 2023

%\cite{Chen:2022wkh}
\bibitem{Chen:2022wkh}
K.~Chen, Z.~Y.~Lin, and S.~L.~Zhu,
Comparison between the $P_{\psi}^N$ and $P_{\psi s}^\Lambda$ systems,
\href{https://journals.aps.org/prd/abstract/10.1103/PhysRevD.106.116017}{Phys. Rev. D \textbf{106}, 116017 (2022)}.
%doi:10.1103/PhysRevD.106.116017
%[arXiv:2211.05558 [hep-ph]].
%7 citations counted in INSPIRE as of 03 Jul 2023

%\cite{Karliner:2022erb}
\bibitem{Karliner:2022erb}
M.~Karliner and J.~R.~Rosner,
New strange pentaquarks,
\href{https://journals.aps.org/prd/abstract/10.1103/PhysRevD.106.036024}{Phys. Rev. D \textbf{106}, 036024 (2022)}.
%0 citations counted in INSPIRE as of 21 Jul 2022

%\cite{Wang:2022mxy}
\bibitem{Wang:2022mxy}
F.~L.~Wang and X.~Liu,
Emergence of molecular-type characteristic spectrum of hidden-charm pentaquark with strangeness embodied in the $P_{\psi s}^\Lambda(4338)$ and $P_{cs}(4459)$,
\href{https://www.sciencedirect.com/science/article/pii/S0370269322007171?via\%3Dihub}{Phys. Lett. B \textbf{835}, 137583 (2022)}.
%doi:10.1016/j.physletb.2022.137583
%[arXiv:2207.10493 [hep-ph]].
%20 citations counted in INSPIRE as of 03 Jul 2023

%\cite{Peng:2019wys}
\bibitem{Peng:2019wys}
F.~Z.~Peng, M.~Z.~Liu, Y.~W.~Pan, M.~S\'anchez S\'anchez, and M.~Pavon Valderrama,
Five-flavor pentaquarks and other light- and heavy-flavor symmetry partners of the LHCb hidden-charm pentaquarks,
\href{https://www.sciencedirect.com/science/article/pii/S0550321322002875?via\%3Dihub}{Nucl. Phys. B \textbf{983}, 115936 (2022)}.
%doi:10.1016/j.nuclphysb.2022.115936
%[arXiv:1907.05322 [hep-ph]].
%15 citations counted in INSPIRE as of 03 Jul 2023

%\cite{Xiao:2022csb}
\bibitem{Xiao:2022csb}
C.~W.~Xiao, J.~Nieves, E.~Oset, J.~J.~Wu, and B.~S.~Zou,
Is $P_{cs}(4459)$ one state or two?,
\href{https://rmf.smf.mx/ojs/index.php/rmf-s/article/view/6298}{Rev. Mex. Fis. Suppl. \textbf{3}, 0308045 (2022)}.
%doi:10.31349/SuplRevMexFis.3.0308045
%0 citations counted in INSPIRE as of 03 Jul 2023

%\cite{Wang:2022gfb}
\bibitem{Wang:2022gfb}
X.~W.~Wang and Z.~G.~Wang,
Study of isospin eigenstates of the pentaquark molecular states with strangeness,
\href{https://www.worldscientific.com/doi/10.1142/S0217751X22501895}{Int. J. Mod. Phys. A \textbf{37}, 2250189 (2022)}.
%doi:10.1142/S0217751X22501895
%[arXiv:2205.02530 [hep-ph]].
%7 citations counted in INSPIRE as of 03 Jul 2023

%\cite{Clymton:2022qlr}
\bibitem{Clymton:2022qlr}
S.~Clymton, H.~J.~Kim, and H.~C.~Kim,
The effect of hidden-charm strange pentaquarks $p_{cs}$ on the $K^- p \to J/\psi \Lambda$~ reaction,
\href{https://rmf.smf.mx/ojs/index.php/rmf-s/article/view/6209}{Rev. Mex. Fis. Suppl. \textbf{3}, 0308040 (2022)}.
%doi:10.31349/SuplRevMexFis.3.0308040
%0 citations counted in INSPIRE as of 03 Jul 2023

%\cite{Ferretti:2021zis}
\bibitem{Ferretti:2021zis}
J.~Ferretti and E.~Santopinto,
The new $P_{\rm cs}(4459)$, $Z_{\rm cs}(3985)$, $Z_{\rm cs}(4000)$ and $Z_{\rm cs}(4220)$ and the possible emergence of flavor pentaquark octets and tetraquark nonets,
\href{https://www.sciencedirect.com/science/article/pii/S2095927322001335?via\%3Dihub}{Sci. Bull. \textbf{67}, 1209 (2022)}.
%doi:10.1016/j.scib.2022.04.010
%[arXiv:2111.08650 [hep-ph]].
%13 citations counted in INSPIRE as of 03 Jul 2023

%\cite{Giachino:2022pws}
\bibitem{Giachino:2022pws}
A.~Giachino, A.~Hosaka, E.~Santopinto, S.~Takeuchi, M.~Takizawa, and Y.~Yamaguchi,
Rich structure of the hidden-charm pentaquarks near threshold regions,
\href{https://journals.aps.org/prd/abstract/10.1103/PhysRevD.108.074012}{Phys. Rev. D \textbf{108}, 074012 (2023)}.
%5 citations counted in INSPIRE as of 03 Jul 2023

%\cite{Nakamura:2022jpd}
\bibitem{Nakamura:2022jpd}
S.~X.~Nakamura and J.~J.~Wu,
Pole determination of $P_{\psi s}^\Lambda(4338)$ and possible $P_{\psi s}^\Lambda(4255)$ in $B^-\to J/\psi\Lambda\bar{p}$,
\href{https://journals.aps.org/prd/abstract/10.1103/PhysRevD.108.L011501}{Phys. Rev. D \textbf{108}, L011501 (2023)}.
%10 citations counted in INSPIRE as of 03 Jul 2023

\bibitem{Yan:2022wuz}
M.~J.~Yan, F.~Z.~Peng, M.~S\'anchez S\'anchez, and M.~Pavon Valderrama,
$P_{\psi s}^{\Lambda}(4338)$ pentaquark and its partners in the molecular picture,
\href{https://journals.aps.org/prd/abstract/10.1103/PhysRevD.107.074025}{Phys. Rev. D \textbf{107}, 074025 (2023)}.
%doi:10.1103/PhysRevD.107.074025
%[arXiv:2207.11144 [hep-ph]].
%17 citations counted in INSPIRE as of 03 Jul 2023

%\cite{Azizi:2023foj}
\bibitem{Azizi:2023foj}
K.~Azizi, Y.~Sarac, and H.~Sundu,
Investigation of the strange pentaquark candidate $P_{\psi s}^{\Lambda}(4338){}^0$ recently observed by LHCb,
\href{https://journals.aps.org/prd/abstract/10.1103/PhysRevD.108.074010}{Phys. Rev. D \textbf{108}, 074010 (2023)}.
%0 citations counted in INSPIRE as of 03 Jul 2023

%\cite{Feijoo:2022rxf}
\bibitem{Feijoo:2022rxf}
A.~Feijoo, W.~F.~Wang, C.~W.~Xiao, J.~J.~Wu, E.~Oset, J.~Nieves, and B.~S.~Zou,
A new look at the $P_{cs}$ states from a molecular perspective,
\href{https://www.sciencedirect.com/science/article/pii/S0370269323000941?via\%3Dihub}{Phys. Lett. B \textbf{839}, 137760 (2023)}.
%doi:10.1016/j.physletb.2023.137760
%[arXiv:2212.12223 [hep-ph]].
%6 citations counted in INSPIRE as of 03 Jul 2023

%\cite{Zhu:2022wpi}
\bibitem{Zhu:2022wpi}
J.~T.~Zhu, S.~Y.~Kong, and J.~He,
$P_{\psi s}^\Lambda(4459)$ and $P_{\psi s}^\Lambda(4338)$ as molecular states in $J/\psi \Lambda$ invariant mass spectra,
\href{https://journals.aps.org/prd/abstract/10.1103/PhysRevD.107.034029}{Phys. Rev. D \textbf{107}, 034029 (2023)}.
%doi:10.1103/PhysRevD.107.034029
%[arXiv:2211.06232 [hep-ph]].
%6 citations counted in INSPIRE as of 03 Jul 2023

%\cite{Ortega:2022uyu}
\bibitem{Ortega:2022uyu}
P.~G.~Ortega, D.~R.~Entem, and F.~Fernandez,
Strange hidden-charm $P_{\psi s}^\Lambda(4459)$ and $P_{\psi s}^\Lambda(4338)$ pentaquarks and additional $P_{\psi s}^{\Lambda}$, $P_{\psi s}^{\Sigma}$ and $P_{\psi ss}^{N}$ candidates in a quark model approach,
\href{https://www.sciencedirect.com/science/article/pii/S0370269323000813?via\%3Dihub}{Phys. Lett. B \textbf{838}, 137747 (2023)}.
%doi:10.1016/j.physletb.2023.137747
%[arXiv:2210.04465 [hep-ph]].
%8 citations counted in INSPIRE as of 03 Jul 2023

%\cite{Meng:2022wgl}
\bibitem{Meng:2022wgl}
L.~Meng, B.~Wang, and S.~L.~Zhu,
Double thresholds distort the line shapes of the $P_{\psi s}^\Lambda(4338)^0$ resonance,
\href{https://journals.aps.org/prd/abstract/10.1103/PhysRevD.107.014005}{Phys. Rev. D \textbf{107}, 014005 (2023)}.
%doi:10.1103/PhysRevD.107.014005
%[arXiv:2208.03883 [hep-ph]].
%14 citations counted in INSPIRE as of 03 Jul 2023

%\cite{Wang:2022neq}
\bibitem{Wang:2022neq}
X.~W.~Wang and Z.~G.~Wang,
Analysis of $P_{cs}(4338)$ and related pentaquark molecular states via QCD sum rules,
\href{https://iopscience.iop.org/article/10.1088/1674-1137/ac9aab}{Chin. Phys. C \textbf{47}, 013109 (2023)}.
%doi:10.1088/1674-1137/ac9aab
%[arXiv:2207.06060 [hep-ph]].
%14 citations counted in INSPIRE as of 03 Jul 2023

%\cite{Yang:2022ezl}
\bibitem{Yang:2022ezl}
Z.~Y.~Yang, F.~Z.~Peng, M.~J.~Yan, M.~S\'anchez S\'anchez, and M.~Pavon Valderrama,
Molecular $P_{\psi}$ pentaquarks from light-meson exchange saturation,
\href{https://journals.aps.org/prd/abstract/10.1103/PhysRevD.111.014012}{Phys. Rev. D \textbf{111}, 014012 (2025)}.
%8 citations counted in INSPIRE as of 03 Jul 2023

%\cite{D0:2016mwd}
\bibitem{D0:2016mwd}
V.~M.~Abazov \textit{et al.} [D0],
Evidence for a $B_s^0 \pi^\pm$ state,
\href{https://link.aps.org/doi/10.1103/PhysRevLett.117.022003}{Phys. Rev. Lett. \textbf{117}, 022003 (2016)}.
%doi:10.1103/PhysRevLett.117.022003
%[arXiv:1602.07588 [hep-ex]].
%264 citations counted in INSPIRE as of 13 Mar 2026

%\cite{LHCb:2016dxl}
\bibitem{LHCb:2016dxl}
R.~Aaij \textit{et al.} [LHCb],
Search for structure in the $B_s^0\pi^\pm$ invariant mass spectrum,
\href{https://journals.aps.org/prl/abstract/10.1103/PhysRevLett.117.152003}{Phys. Rev. Lett. \textbf{117}, 152003 (2016)}.
%doi:10.1103/PhysRevLett.117.152003
%[arXiv:1608.00435 [hep-ex]].
%146 citations counted in INSPIRE as of 13 Mar 2026

%\cite{CDF:2017dwr}
\bibitem{CDF:2017dwr}
T.~Aaltonen \textit{et al.} [CDF],
A search for the exotic meson $X(5568)$ with the collider detector at Fermilab,
\href{https://journals.aps.org/prl/abstract/10.1103/PhysRevLett.120.202006}{Phys. Rev. Lett. \textbf{120}, 202006 (2018)}.
%doi:10.1103/PhysRevLett.120.202006
%[arXiv:1712.09620 [hep-ex]].
%64 citations counted in INSPIRE as of 13 Mar 2026

%\cite{CMS:2017hfy}
\bibitem{CMS:2017hfy}
A.~M.~Sirunyan \textit{et al.} [CMS],
Search for the X(5568) state decaying into $B_s^0\pi^\pm$ in proton-proton collisions at $\sqrt{s}=$8 TeV,
\href{https://journals.aps.org/prl/abstract/10.1103/PhysRevLett.120.202005}{Phys. Rev. Lett. \textbf{120}, 202005 (2018)}.
%doi:10.1103/PhysRevLett.120.202005
%[arXiv:1712.06144 [hep-ex]].
%87 citations counted in INSPIRE as of 13 Mar 2026

%\cite{ATLAS:2018udc}
\bibitem{ATLAS:2018udc}
M.~Aaboud \textit{et al.} [ATLAS],
Search for a structure in the $B^0_s \pi^\pm$ invariant mass spectrum with the ATLAS experiment,
\href{https://journals.aps.org/prl/abstract/10.1103/PhysRevLett.120.202007}{Phys. Rev. Lett. \textbf{120}, 202007 (2018)}.
%doi:10.1103/PhysRevLett.120.202007
%[arXiv:1802.01840 [hep-ex]].
%65 citations counted in INSPIRE as of 13 Mar 2026

%\cite{LHCb:2020bls}
\bibitem{LHCb:2020bls}
R.~Aaij \textit{et al.} [LHCb],
A model-independent study of resonant structure in $B^+\to D^+D^-K^+$ decays,
\href{https://journals.aps.org/prl/abstract/10.1103/PhysRevLett.125.242001}{Phys. Rev. Lett. \textbf{125}, 242001 (2020)}.
%doi:10.1103/PhysRevLett.125.242001
%[arXiv:2009.00025 [hep-ex]].
%243 citations counted in INSPIRE as of 29 Sep 2025

%\cite{LHCb:2020pxc}
\bibitem{LHCb:2020pxc}
R.~Aaij \textit{et al.} [LHCb],
Amplitude analysis of the $B^+\to D^+D^-K^+$ decay,
\href{https://journals.aps.org/prd/abstract/10.1103/PhysRevD.102.112003}{Phys. Rev. D \textbf{102}, 112003 (2020)}.
%doi:10.1103/PhysRevD.102.112003
%[arXiv:2009.00026 [hep-ex]].
%226 citations counted in INSPIRE as of 11 Jun 2024

%\cite{LHCb:2024vfz}
\bibitem{LHCb:2024vfz}
R.~Aaij \textit{et al.} [LHCb],
Observation of new charmonium or charmoniumlike states in $B^+ \to D^{*\pm}D^{\mp}K^+$ decays,
\href{https://journals.aps.org/prl/abstract/10.1103/PhysRevLett.133.131902}{Phys. Rev. Lett. \textbf{133}, 131902 (2024)}.
%doi:10.1103/PhysRevLett.133.131902
%[arXiv:2406.03156 [hep-ex]].
%31 citations counted in INSPIRE as of 29 Sep 2025

%\cite{Molina:2020hde}
\bibitem{Molina:2020hde}
R.~Molina and E.~Oset,
Molecular picture for the $X_0(2866)$ as a $D^* \bar{K}^*$ $J^P=0^+$ state and related $1^+,2^+$ states,
\href{https://linkinghub.elsevier.com/retrieve/pii/S0370269320306730}{Phys. Lett. B \textbf{811}, 135870 (2020); \textbf{837}, 137645(E) (2023)}.
%doi:10.1016/j.physletb.2020.135870
%[arXiv:2008.11171 [hep-ph]].
%64 citations counted in INSPIRE as of 11 Jun 2024

%\cite{Liu:2020nil}
\bibitem{Liu:2020nil}
M.~Z.~Liu, J.~J.~Xie, and L.~S.~Geng,
$X_0(2866)$ as a $D^*\bar{K}^*$ molecular state,
\href{https://journals.aps.org/prd/pdf/10.1103/PhysRevD.102.091502}{Phys. Rev. D \textbf{102}, 091502 (2020)}.
%doi:10.1103/PhysRevD.102.091502
%[arXiv:2008.07389 [hep-ph]].
%90 citations counted in INSPIRE as of 06 Oct 2024

%\cite{Chen:2020aos}
\bibitem{Chen:2020aos}
H.~X.~Chen, W.~Chen, R.~R.~Dong, and N.~Su,
$X_0$(2900) and $X_1$(2900): Hadronic molecules or compact tetraquarks,
\href{https://iopscience.iop.org/article/10.1088/0256-307X/37/10/101201}{Chin. Phys. Lett. \textbf{37}, 101201 (2020)}.
%74 citations counted in INSPIRE as of 11 Jun 2024

%\cite{Hu:2020mxp}
\bibitem{Hu:2020mxp}
M.~W.~Hu, X.~Y.~Lao, P.~Ling, and Q.~Wang,
$X_0$(2900) and its heavy quark spin partners in molecular picture,
\href{https://iopscience.iop.org/article/10.1088/1674-1137/abcfaa}{Chin. Phys. C \textbf{45}, 021003 (2021)}.
%doi:10.1088/1674-1137/abcfaa
%[arXiv:2008.06894 [hep-ph]].
%65 citations counted in INSPIRE as of 30 Sep 2025

%\cite{Kong:2021ohg}
\bibitem{Kong:2021ohg}
S.~Y.~Kong, J.~T.~Zhu, D.~Song, and J.~He,
Heavy-strange meson molecules and possible candidates $D_{s0}^{*}(2317)$, $D_{s1}(2460)$, and $X_0(2900)$,
\href{https://journals.aps.org/prd/abstract/10.1103/PhysRevD.104.094012}{Phys. Rev. D \textbf{104}, 094012 (2021)}.
%doi:10.1103/PhysRevD.104.094012
%[arXiv:2106.07272 [hep-ph]].
%28 citations counted in INSPIRE as of 11 Jun 2024

%\cite{Agaev:2020nrc}
\bibitem{Agaev:2020nrc}
S.~S.~Agaev, K.~Azizi, and H.~Sundu,
New scalar resonance $X_0(2900)$ as a molecule: Mass and width,
\href{https://iopscience.iop.org/article/10.1088/1361-6471/ac0b31}{J. Phys. G \textbf{48}, 085012 (2021)}.
%doi:10.1088/1361-6471/ac0b31
%[arXiv:2008.13027 [hep-ph]].
%47 citations counted in INSPIRE as of 11 Jun 2024

%\cite{He:2020btl}
\bibitem{He:2020btl}
J.~He and D.~Y.~Chen,
Molecular picture for $X_0(2900)$ and $X_1(2900)$,
\href{https://iopscience.iop.org/article/10.1088/1674-1137/abeda8}{Chin. Phys. C \textbf{45}, 063102 (2021)}.
%doi:10.1088/1674-1137/abeda8
%[arXiv:2008.07782 [hep-ph]].
%46 citations counted in INSPIRE as of 11 Jun 2024

%\cite{Wang:2021lwy}
\bibitem{Wang:2021lwy}
B.~Wang and S.~L.~Zhu,
How to understand the $X(2900)$?,
\href{https://link.springer.com/article/10.1140/epjc/s10052-022-10396-9}{Eur. Phys. J. C \textbf{82}, 419 (2022)}.
%doi:10.1140/epjc/s10052-022-10396-9
%[arXiv:2107.09275 [hep-ph]].
%24 citations counted in INSPIRE as of 11 Jun 2024


%\cite{Dai:2022htx}
\bibitem{Dai:2022htx}
L.~R.~Dai, R.~Molina, and E.~Oset,
Looking for the exotic $X_0(2866)$ and its $J^P=1^+$ partner in the $\bar B^0 \to D^{(*)+}K^-K^{(*)0}$ reactions,
\href{https://journals.aps.org/prd/abstract/10.1103/PhysRevD.105.096022}{Phys. Rev. D \textbf{105}, 096022 (2022)}.
%doi:10.1103/PhysRevD.105.096022
%[arXiv:2202.11973 [hep-ph]].
%15 citations counted in INSPIRE as of 30 Sep 2025

%\cite{Chen:2023syh}
\bibitem{Chen:2023syh}
Y.~K.~Chen, W.~L.~Wu, L.~Meng, and S.~L.~Zhu,
Unified description of the $Qs \bar q \bar q$ molecular bound states, molecular resonances, and compact tetraquark states in the quark potential model,
\href{https://journals.aps.org/prd/abstract/10.1103/PhysRevD.109.014010}{Phys. Rev. D \textbf{109}, 014010 (2024)}.
%doi:10.1103/PhysRevD.109.014010
%[arXiv:2310.14597 [hep-ph]].
%12 citations counted in INSPIRE as of 30 Sep 2025

%\cite{Ding:2024dif}
\bibitem{Ding:2024dif}
Z.~M.~Ding, Q.~Huang, and J.~He,
$X_0(2900)$ and $\chi _{c0}(3930)$ in process $B^+\rightarrow D^+ D^- K^+$,
\href{https://link.springer.com/article/10.1140/epjc/s10052-024-13214-6}{Eur. Phys. J. C \textbf{84}, 822 (2024)}.
%doi:10.1140/epjc/s10052-024-13214-6
%[arXiv:2407.13503 [hep-ph]].
%5 citations counted in INSPIRE as of 30 Sep 2025

%\cite{Ding:2025uhh}
\bibitem{Ding:2025uhh}
Z.~M.~Ding, Q.~Huang, and J.~He,
Roles of $\bar{D}^{*}K^{*}$ and $D^{*}\bar{D}$ molecular states in decay $B^+ \to D^{*+} D^{-} K^{+}$,
\href{https://link.springer.com/article/10.1140/epjc/s10052-025-14882-8}{Eur. Phys. J. C \textbf{85}, 1133 (2025)}.
%doi:10.1140/epjc/s10052-025-14882-8
%[arXiv:2508.12686 [hep-ph]].
%3 citations counted in INSPIRE as of 24 Mar 2026

%\cite{Dong:2020rgs}
\bibitem{Dong:2020rgs}
X.~K.~Dong and B.~S.~Zou,
Prediction of possible $DK_1$ bound states,
\href{https://link.springer.com/article/10.1140/epja/s10050-021-00442-7}{Eur. Phys. J. A \textbf{57}, 139 (2021)}.
%doi:10.1140/epja/s10050-021-00442-7
%[arXiv:2009.11619 [hep-ph]].
%20 citations counted in INSPIRE as of 30 Sep 2025

%\cite{Qi:2021iyv}
\bibitem{Qi:2021iyv}
J.~J.~Qi, Z.~Y.~Wang, Z.~F.~Zhang, and X.~H.~Guo,
Studying the ${\bar{D}}_1K$ molecule in the Bethe\textendash{}Salpeter equation approach,
\href{https://link.springer.com/article/10.1140/epjc/s10052-021-09422-z}{Eur. Phys. J. C \textbf{81}, 639 (2021)}.
%doi:10.1140/epjc/s10052-021-09422-z
%[arXiv:2101.06688 [hep-ph]].
%11 citations counted in INSPIRE as of 11 Jun 2024

%\cite{Chen:2021tad}
\bibitem{Chen:2021tad}
H.~Chen, H.~R.~Qi, and H.~Q.~Zheng,
$X_1(2900)$ as a $\bar{D}_1 K$ molecule,
\href{https://link.springer.com/article/10.1140/epjc/s10052-021-09603-w}{Eur. Phys. J. C \textbf{81}, 812 (2021)}.
%doi:10.1140/epjc/s10052-021-09603-w
%[arXiv:2108.02387 [hep-ph]].
%16 citations counted in INSPIRE as of 11 Jun 2024

%\cite{Wang:2025wpc}
\bibitem{Wang:2025wpc}
F.~L.~Wang, S.~Q.~Luo, and X.~Liu,
Predicting charmed-strange molecular tetraquarks with $K^{(*)}$ and $T$-doublet (anti-)charmed meson,
\href{https://arxiv.org/abs/2510.17244}{arXiv:2510.17244}.
%0 citations counted in INSPIRE as of 13 Mar 2026

%\cite{Shen:2022rpn}
\bibitem{Shen:2022rpn}
C.~W.~Shen and U.~G.~Mei{\ss}ner,
Prediction of five-flavored pentaquarks,
\href{https://www.sciencedirect.com/science/article/pii/S0370269322003318?via%3Dihub}{Phys. Lett. B \textbf{831}, 137197 (2022)}.
%doi:10.1016/j.physletb.2022.137197
%[arXiv:2203.09804 [hep-ph]].
%7 citations counted in INSPIRE as of 13 Mar 2026

%\cite{Lin:2023iww}
\bibitem{Lin:2023iww}
J.~X.~Lin, H.~X.~Chen, W.~H.~Liang, W.~Y.~Liu, and D.~Zhou,
Molecular pentaquark states with open charm and bottom flavors,
\href{https://link.springer.com/article/10.1140/epja/s10050-024-01240-7}{Eur. Phys. J. A \textbf{60}, 15 (2024)}.
%doi:10.1140/epja/s10050-024-01240-7
%[arXiv:2308.01007 [hep-ph]].
%9 citations counted in INSPIRE as of 13 Mar 2026

%\cite{Chen:2026cag}
\bibitem{Chen:2026cag}
X.~Chen and L.~Ma,
Recoil corrections to pentaquark molecules with an SU(3) anti-triplet heavy baryon,
\href{https://arxiv.org/pdf/2603.21572v1}{arXiv:2603.21572}.
%0 citations counted in INSPIRE as of 24 Mar 2026

%\cite{ParticleDataGroup:2024cfk}
\bibitem{ParticleDataGroup:2024cfk}
S.~Navas \textit{et al.} [Particle Data Group],
Review of particle physics,
\href{https://journals.aps.org/prd/abstract/10.1103/PhysRevD.110.030001}{Phys. Rev. D \textbf{110}, 030001 (2024)}.
%doi:10.1103/PhysRevD.110.030001
%2709 citations counted in INSPIRE as of 29 Sep 2025

%\cite{Berestetskii:1982qgu}
\bibitem{Berestetskii:1982qgu}
 V. B. Berestetsky, E. M. Lifshitz, and L. P. Pitaevsky,
 Quantum Electrodynamics,
 Pergamon Press, 1982, ISBN 978-0-7506-3371-0.

%\cite{Wise:1992hn}
\bibitem{Wise:1992hn}
  M.~B.~Wise,
  Chiral perturbation theory for hadrons containing a heavy quark,
   \href{https://journals.aps.org/prd/abstract/10.1103/PhysRevD.45.R2188}{Phys.\ Rev.\ D {\bf 45}, R2188 (1992)}.
  %doi:10.1103/PhysRevD.45.R2188
  %%CITATION = doi:10.1103/PhysRevD.45.R2188;%%
  %681 citations counted in INSPIRE as of 27 Dec 2018

%\cite{Casalbuoni:1992gi}
\bibitem{Casalbuoni:1992gi}
  R.~Casalbuoni, A.~Deandrea, N.~Di Bartolomeo, R.~Gatto, F.~Feruglio, and G.~Nardulli,
  Light vector resonances in the effective chiral Lagrangian for heavy mesons,
  \href{https://www.sciencedirect.com/science/article/abs/pii/037026939291189G}{Phys.\ Lett.\ B {\bf 292}, 371 (1992)}.
  %doi:10.1016/0370-2693(92)91189-G
  %[hep-ph/9209248].
  %%CITATION = doi:10.1016/0370-2693(92)91189-G;%%
  %95 citations counted in INSPIRE as of 27 Dec 2018

%\cite{Yan:1992gz}
\bibitem{Yan:1992gz}
  T.~M.~Yan, H.~Y.~Cheng, C.~Y.~Cheung, G.~L.~Lin, Y.~C.~Lin, and H.~L.~Yu,
  Heavy quark symmetry and chiral dynamics,
  \href{https://journals.aps.org/prd/abstract/10.1103/PhysRevD.46.1148}{Phys.\ Rev.\ D {\bf 46}, 1148 (1992)};
   \href{https://journals.aps.org/prd/abstract/10.1103/PhysRevD.55.5851}{[Phys.\ Rev.\ D {\bf 55}, 5851E (1997)]}.
  %doi:10.1103/PhysRevD.46.1148, 10.1103/PhysRevD.55.5851
  %%CITATION = doi:10.1103/PhysRevD.46.1148, 10.1103/PhysRevD.55.5851;%%
  %604 citations counted in INSPIRE as of 27 Dec 2018

%\cite{Casalbuoni:1996pg}
\bibitem{Casalbuoni:1996pg}
  R.~Casalbuoni, A.~Deandrea, N.~Di Bartolomeo, R.~Gatto, F.~Feruglio, and G.~Nardulli,
  Phenomenology of heavy meson chiral Lagrangians,
  \href{https://www.sciencedirect.com/science/article/pii/S0370157396000270}{Phys.\ Rep.\  {\bf 281}, 145 (1997)}.
  %doi:10.1016/S0370-1573(96)00027-0
  %[hep-ph/9605342].
  %%CITATION = doi:10.1016/S0370-1573(96)00027-0;%%
  %484 citations counted in INSPIRE as of 27 Dec 2018

%\cite{Bando:1987br}
\bibitem{Bando:1987br}
  M.~Bando, T.~Kugo, and K.~Yamawaki,
  Nonlinear Realization and Hidden Local Symmetries,
  \href{https://www.sciencedirect.com/science/article/abs/pii/0370157388900191?via\%3Dihub}{Phys.\ Rept.\  {\bf 164}, 217 (1988)}.
  %%doi:10.1016/0370-1573(88)90019-1
  %%CITATION = doi:10.1016/0370-1573(88)90019-1;%%
  %1254 citations counted in INSPIRE as of 24 Jan 2021

%\cite{Harada:2003jx}
\bibitem{Harada:2003jx}
  M.~Harada and K.~Yamawaki,
  Hidden local symmetry at loop: A New perspective of composite gauge boson and chiral phase transition,
  \href{https://www.sciencedirect.com/science/article/abs/pii/S037015730300139X?via\%3Dihub}{Phys.\ Rept.\  {\bf 381}, 1 (2003)}.
  %%doi:10.1016/S0370-1573(03)00139-X
  %%[hep-ph/0302103].
  %%CITATION = doi:10.1016/S0370-1573(03)00139-X;%%
  %540 citations counted in INSPIRE as of 24 Jan 2021

%\cite{Chen:2017xat}
\bibitem{Chen:2017xat}
  R.~Chen, A.~Hosaka, and X.~Liu,
  Searching for possible $\Omega_c$-like molecular states from meson-baryon interaction,
  \href{https://journals.aps.org/prd/abstract/10.1103/PhysRevD.97.036016}{Phys.\ Rev.\ D {\bf 97}, 036016 (2018)}.
  %doi:10.1103/PhysRevD.97.036016
  %[arXiv:1711.07650 [hep-ph]].
  %%CITATION = doi:10.1103/PhysRevD.97.036016;%%
  %12 citations counted in INSPIRE as of 20 Apr 2019

%\cite{Ding:2008gr}
\bibitem{Ding:2008gr}
  G.~J.~Ding,
  Are $Y(4260)$ and {\rm$Z_2^{+}$(4250)} $D_1D$ or $D_0 D^{*}$ hadronic molecules?
  \href{https://journals.aps.org/prd/abstract/10.1103/PhysRevD.79.014001}{Phys.\ Rev.\ D {\bf 79}, 014001 (2009)}.
  %doi:10.1103/PhysRevD.79.014001
  %[arXiv:0809.4818 [hep-ph]].
  %%CITATION = doi:10.1103/PhysRevD.79.014001;%%
  %124 citations counted in INSPIRE as of 27 Dec 2018

%\cite{Chen:2018pzd}
\bibitem{Chen:2018pzd}
  R.~Chen, F.~L.~Wang, A.~Hosaka, and X.~Liu,
  Exotic triple-charm deuteronlike hexaquarks,
  \href{https://journals.aps.org/prd/abstract/10.1103/PhysRevD.97.114011}{Phys.\ Rev.\ D {\bf 97}, 114011 (2018)}.
  %%doi:10.1103/PhysRevD.97.114011
  %%[arXiv:1804.02961 [hep-ph]].
  %%CITATION = doi:10.1103/PhysRevD.97.114011;%%
  %13 citations counted in INSPIRE as of 23 Apr 2021

%\cite{Chen:2019asm}
\bibitem{Chen:2019asm}
  R.~Chen, Z.~F.~Sun, X.~Liu, and S.~L.~Zhu,
  Strong LHCb evidence supporting the existence of the hidden-charm molecular pentaquarks,
  \href{https://journals.aps.org/prd/abstract/10.1103/PhysRevD.100.011502}{Phys.\ Rev.\ D {\bf 100}, 011502 (2019)}.
  %%doi:10.1103/PhysRevD.100.011502
  %%[arXiv:1903.11013 [hep-ph]].
  %%CITATION = doi:10.1103/PhysRevD.100.011502;%%
  %38 citations counted in INSPIRE as of 22 Jul 2019

 %\cite{Wang:2020dya}
\bibitem{Wang:2020dya}
F.~L.~Wang and X.~Liu,
Exotic double-charm molecular states with hidden or open strangeness and around $4.5\sim 4.7$ GeV,
\href{https://journals.aps.org/prd/abstract/10.1103/PhysRevD.102.094006}{Phys. Rev. D \textbf{102}, 094006 (2020)}.
%doi:10.1103/PhysRevD.102.094006
%[arXiv:2008.13484 [hep-ph]].
%0 citations counted in INSPIRE as of 11 Nov 2020

%\cite{Wang:2021yld}
\bibitem{Wang:2021yld}
F.~L.~Wang and X.~Liu,
Investigating new type of doubly charmed molecular tetraquarks composed of charmed mesons in the $H$ and $T$ doublets,
\href{https://journals.aps.org/prd/abstract/10.1103/PhysRevD.104.094030}{Phys. Rev. D \textbf{104}, 094030 (2021)}.
%doi:10.1103/PhysRevD.104.094030
%[arXiv:2108.09925 [hep-ph]].
%14 citations counted in INSPIRE as of 03 Jul 2023

%\cite{Wang:2021aql}
\bibitem{Wang:2021aql}
F.~L.~Wang, X.~D.~Yang, R.~Chen, and X.~Liu,
Correlation of the hidden-charm molecular tetraquarks and the charmoniumlike structures existing in the $B\to XYZ+K$ process,
\href{https://journals.aps.org/prd/abstract/10.1103/PhysRevD.104.094010}{Phys. Rev. D \textbf{104}, 094010 (2021)}.
%doi:10.1103/PhysRevD.104.094010
%[arXiv:2103.04698 [hep-ph]].
%15 citations counted in INSPIRE as of 03 Jul 2023

%\cite{Wang:2020bjt}
\bibitem{Wang:2020bjt}
F.~L.~Wang, R.~Chen, and X.~Liu,
Prediction of hidden-charm pentaquarks with double strangeness,
\href{https://journals.aps.org/prd/abstract/10.1103/PhysRevD.103.034014}{Phys. Rev. D \textbf{103}, 034014 (2021)}.
%doi:10.1103/PhysRevD.103.034014
%[arXiv:2011.14296 [hep-ph]].
%3 citations counted in INSPIRE as of 12 Mar 2021

%\cite{Wang:2021hql}
\bibitem{Wang:2021hql}
F.~L.~Wang, X.~D.~Yang, R.~Chen, and X.~Liu,
Hidden-charm pentaquarks with triple strangeness due to the $\Omega_{c}^{(*)}\bar{D}_s^{(*)}$ interactions,
\href{https://journals.aps.org/prd/abstract/10.1103/PhysRevD.103.054025}{Phys. Rev. D \textbf{103}, 054025 (2021)}.
%doi:10.1103/PhysRevD.103.054025
%[arXiv:2101.11200 [hep-ph]].
%2 citations counted in INSPIRE as of 31 Mar 2021

%\cite{Yang:2021sue}
\bibitem{Yang:2021sue}
X.~D.~Yang, F.~L.~Wang, Z.~W.~Liu, and X.~Liu,
Newly observed X(4630): a new charmoniumlike molecule,
\href{https://link.springer.com/article/10.1140/epjc/s10052-021-09606-7}{Eur. Phys. J. C \textbf{81}, 807 (2021)}.
%doi:10.1140/epjc/s10052-021-09606-7
%[arXiv:2103.03127 [hep-ph]].
%10 citations counted in INSPIRE as of 03 Jul 2023

%\cite{Wang:2021ajy}
\bibitem{Wang:2021ajy}
F.~L.~Wang, R.~Chen, and X.~Liu,
A new group of doubly charmed molecule with $T$-doublet charmed meson pair,
\href{https://www.sciencedirect.com/science/article/pii/S0370269322006360?via\%3Dihub}{Phys. Lett. B \textbf{835}, 137502 (2022)}.
%doi:10.1016/j.physletb.2022.137502
%[arXiv:2111.00208 [hep-ph]].
%9 citations counted in INSPIRE as of 03 Jul 2023

%\cite{Wang:2023ftp}
\bibitem{Wang:2023ftp}
F.~L.~Wang and X.~Liu,
Higher molecular $P_{\psi s}^{\Lambda/\Sigma}$ pentaquarks arising from the $\Xi_c^{(\prime,*)}\bar D_1/\Xi_c^{(\prime,*)}\bar D_2^*$ interactions,
\href{https://journals.aps.org/prd/abstract/10.1103/PhysRevD.108.054028}{Phys. Rev. D \textbf{108}, 054028 (2023)}.
%doi:10.1103/PhysRevD.108.054028
%[arXiv:2307.08276 [hep-ph]].
%2 citations counted in INSPIRE as of 15 Nov 2023

%\cite{Yalikun:2023waw}
\bibitem{Yalikun:2023waw}
N.~Yalikun, X.~K.~Dong, and B.~S.~Zou,
Molecular states in $D_s^{(*)+}\Xi_c^{(',*)}$ systems,
\href{https://iopscience.iop.org/article/10.1088/1674-1137/acf65e}{Chin. Phys. C \textbf{47}, 123101 (2023)}.
%doi:10.1088/1674-1137/acf65e
%[arXiv:2303.03629 [hep-ph]].
%10 citations counted in INSPIRE as of 24 Mar 2026

%\cite{Wang:2023aob}
\bibitem{Wang:2023aob}
F.~L.~Wang and X.~Liu,
New type of doubly charmed molecular pentaquarks containing most strange quarks: Mass spectra, radiative decays, and magnetic moments,
\href{https://journals.aps.org/prd/abstract/10.1103/PhysRevD.108.074022}{Phys. Rev. D \textbf{108}, 074022 (2023)}.
%doi:10.1103/PhysRevD.108.074022
%[arXiv:2308.15255 [hep-ph]].
%0 citations counted in INSPIRE as of 13 Nov 2023

%\cite{Wang:2019nwt}
\bibitem{Wang:2019nwt}
  F.~L.~Wang, R.~Chen, Z.~W.~Liu, and X.~Liu,
  Probing new types of $P_c$ states inspired by the interaction between $S-$wave charmed baryon and anti-charmed meson in a $\bar T$ doublet,
\href{https://journals.aps.org/prc/abstract/10.1103/PhysRevC.101.025201}{Phys.\ Rev.\ C {\bf 101},  025201 (2020)}.
  %%doi:10.1103/PhysRevC.101.025201
  %%[arXiv:1905.03636 [hep-ph]].
  %%CITATION = doi:10.1103/PhysRevC.101.025201;%%
  %7 citations counted in INSPIRE as of 27 Feb 2020

%\cite{Wang:2023ael}
\bibitem{Wang:2023ael}
F.~L.~Wang and X.~Liu,
Surveying the mass spectra and the electromagnetic properties of the $\Xi_c^{(\prime,*)}D^{(*)}$ molecular pentaquarks,
\href{https://journals.aps.org/prd/abstract/10.1103/PhysRevD.109.014043}{Phys. Rev. D \textbf{109}, 014043 (2024)}.
%doi:10.1103/PhysRevD.109.014043
%[arXiv:2311.13968 [hep-ph]].
%20 citations counted in INSPIRE as of 15 Mar 2026

%\cite{Machleidt:1987hj}
\bibitem{Machleidt:1987hj}
R.~Machleidt, K.~Holinde, and C.~Elster,
The Bonn Meson Exchange Model for the Nucleon Nucleon Interaction,
\href{https://www.sciencedirect.com/science/article/abs/pii/S0370157387800029?via\%3Dihub}{Phys. Rept. \textbf{149}, 1-89 (1987)}.
%doi:10.1016/S0370-1573(87)80002-9
%2397 citations counted in INSPIRE as of 28 Oct 2022

%\cite{Epelbaum:2008ga}
\bibitem{Epelbaum:2008ga}
E.~Epelbaum, H.~W.~Hammer, and U.~G.~Meissner,
Modern Theory of Nuclear Forces,
\href{https://journals.aps.org/rmp/abstract/10.1103/RevModPhys.81.1773}{Rev. Mod. Phys. \textbf{81}, 1773-1825 (2009)}.
%doi:10.1103/RevModPhys.81.1773
%[arXiv:0811.1338 [nucl-th]].
%1482 citations counted in INSPIRE as of 28 Oct 2022

%\cite{Esposito:2014rxa}
\bibitem{Esposito:2014rxa}
A.~Esposito, A.~L.~Guerrieri, F.~Piccinini, A.~Pilloni, and A.~D.~Polosa,
Four-Quark Hadrons: an Updated Review,
\href{https://www.worldscientific.com/doi/abs/10.1142/S0217751X15300021}{Int. J. Mod. Phys. A \textbf{30}, 1530002 (2015)}.
%doi:10.1142/S0217751X15300021
%[arXiv:1411.5997 [hep-ph]].
%218 citations counted in INSPIRE as of 28 Oct 2022

%\cite{Tornqvist:1993ng}
\bibitem{Tornqvist:1993ng}
  N.~A.~Tornqvist,
  From the deuteron to deusons, an analysis of deuteron-like meson-meson bound states,
   \href{https://link.springer.com/article/10.1007\%2FBF01413192}{Z.\ Phys.\ C {\bf 61}, 525 (1994)}.
  %doi:10.1007/BF01413192
 % [hep-ph/9310247].
  %%CITATION = doi:10.1007/BF01413192;%%
  %358 citations counted in INSPIRE as of 27 Dec 2018

%\cite{Tornqvist:1993vu}
\bibitem{Tornqvist:1993vu}
  N.~A.~Tornqvist,
  On deusons or deuteron-like meson-meson bound states,
  \href{https://link.springer.com/article/10.1007\%2FBF02734018}{Nuovo Cim. Soc. Ital. Fis.  {\bf 107A}, 2471 (1994)}.
  %doi:10.1007/BF02734018
  %[hep-ph/9310225].
  %%CITATION = doi:10.1007/BF02734018;%%
  %41 citations counted in INSPIRE as of 27 Dec 2018

%\cite{Chen:2017jjn}
\bibitem{Chen:2017jjn}
  R.~Chen, A.~Hosaka, and X.~Liu,
  Prediction of triple-charm molecular pentaquarks,
  \href{https://journals.aps.org/prd/abstract/10.1103/PhysRevD.96.114030}{Phys.\ Rev.\ D {\bf 96}, 114030 (2017)}.
  %%doi:10.1103/PhysRevD.96.114030
  %%[arXiv:1711.09579 [hep-ph]].
  %%CITATION = doi:10.1103/PhysRevD.96.114030;%%
  %11 citations counted in INSPIRE as of 26 Sep 2019

%\cite{Isgur:1989vq}
\bibitem{Isgur:1989vq}
N.~Isgur and M.~B.~Wise,
Weak decays of heavy mesons in the static quark approximation,
\href{https://www.sciencedirect.com/science/article/abs/pii/0370269389905662?via\%3Dihub}{Phys. Lett. B \textbf{232}, 113-117 (1989)}.
%doi:10.1016/0370-2693(89)90566-2
%2511 citations counted in INSPIRE as of 16 Mar 2026

%\cite{Isgur:1990yhj}
\bibitem{Isgur:1990yhj}
N.~Isgur and M.~B.~Wise,
Weak transition form factors between heavy mesons,
\href{https://www.sciencedirect.com/science/article/abs/pii/0370269390912192?via\%3Dihub}{Phys. Lett. B \textbf{237}, 527-530 (1990)}.
%doi:10.1016/0370-2693(90)91219-2
%2047 citations counted in INSPIRE as of 16 Mar 2026


%\cite{Belle:2025pey}
\bibitem{Belle:2025pey}
I.~Adachi \textit{et al.} [Belle and Belle II],
Search for $P_{c \bar c s}(4459)$ and $P_{c \bar c s}(4338)$ in $\Upsilon(1S,2S)$ inclusive decays at Belle,
\href{https://journals.aps.org/prl/abstract/10.1103/pf8m-6j69}{Phys. Rev. Lett. \textbf{135}, 041901 (2025)}.
%doi:10.1103/pf8m-6j69
%[arXiv:2502.09951 [hep-ex]].
%27 citations counted in INSPIRE as of 16 Mar 2026

\end{thebibliography}
\end{document}